\documentclass[11pt,a4paper]{article}
\usepackage[utf8]{inputenc}
\usepackage[T1]{fontenc}
\usepackage{lmodern}
\usepackage[margin=1in]{geometry}
\usepackage{amsmath,amssymb,amsthm,mathtools}
\usepackage{graphicx,booktabs,tabularx,array,longtable}
\usepackage{microtype}
\usepackage[section]{placeins}
\usepackage{xcolor}
\usepackage{xurl}
\usepackage[authoryear,round]{natbib}
\usepackage{hyperref}
\usepackage{caption}
\hypersetup{colorlinks=true,linkcolor=black,citecolor=black,urlcolor=blue!45!black}
\newtheorem{proposition}{Proposition}
\newtheorem{corollary}{Corollary}
\theoremstyle{remark}

\newcolumntype{Y}{>{\raggedright\arraybackslash}X}
\newcommand{\Q}{\dot Q}
\newcommand{\dd}{\mathrm{d}}
\newcommand{\W}{\,\mathrm{W}}

\DeclareMathOperator{\rank}{rank}

\newif\ifshowauthors
\showauthorstrue
\title{Conservation Bounds and Identifiability in\\
Nasal-to-Cerebral Heat Transfer}
\ifshowauthors
\author{William \'{A}lvarez Gaviria$^{1}$ \qquad Juan D. V\'{e}lez$^{2}$\\[0.65em]
\small $^{1}$Universidad de Antioquia \& Cl\'{i}nica ORLANT, Medell\'{i}n, Colombia\\
\small $^{2}$Departamento de Matem\'{a}ticas, Universidad Nacional de Colombia,\\
\small Sede Medell\'{i}n, Colombia}
\hypersetup{pdfauthor={William Alvarez Gaviria and Juan D. Velez}}
\else
\author{}
\hypersetup{pdfauthor={}}
\fi
\hypersetup{pdftitle={Conservation Bounds and Identifiability in Nasal-to-Cerebral Heat Transfer}}
\date{}
\begin{document}
\maketitle

\begin{abstract}
Quantifying a cerebral thermal contribution from nasal heat exchange requires both a defined source budget and an identifiable observation model. We analyze a reduced allocation among direct conduction, regional vascular transfer and systemic redistribution. For the steady model, we derive sharp output bounds, complete single-endpoint coefficient intervals and a conditional threshold requiring regional transfer. Repeated conditions with fixed cerebral flow fraction admit a two-parameter linear representation and an exact convex compatibility region for bounded endpoint errors. When measured flow fractions vary, a constrained three-variable representation and a two-condition design identify the effective coefficients under explicit invariance assumptions. We then introduce a nonnegative residual-energy store. Its integrated conservation law replaces the instantaneous source bound, and its stationary limit recovers the original allocation. Time-resolved outputs can additionally identify storage rates when the corresponding integral design has rank two. Reproducible synthetic examples illustrate identification, degeneracy and delayed responses without estimating human cooling capacity. Published nasal, head and whole-body thermal models provide complementary forward descriptions but do not independently supply the matched source and cerebral-power endpoint required here. The results define compatibility and measurement requirements within the stated model; they do not establish a positive natural human cooling rate or therapeutic efficacy.
\end{abstract}

\noindent\textbf{Keywords:} nasal heat exchange; bioheat transfer; conservation of energy; structural identifiability; inverse problems; thermal storage; thermoregulation

\section{Introduction}
The nasal cavity exchanges heat and water with respiratory air. Inspiration and expiration contribute with different signs to the net transfer, and heat recovered during expiration must be included in any cycle-averaged budget \citep{Varene1986,Na2020}. Establishing this exchange is different from quantifying its effect on the brain. A cooled nasal surface, a venous connection and a decrease in a local intracranial temperature are distinct observations. Relating them requires a source definition, a transport model and an observable cerebral endpoint.

This distinction matters because cerebral temperature depends on arterial temperature, perfusion, metabolic heat production and exchange with surrounding tissues \citep{Nybo2002,Sukstanskii2004,Sukstanskii2006}. Several mechanisms could therefore contribute to a response during a nasal intervention. Conductive transfer may affect tissue near the skull base; cooled venous blood could participate in regional exchange; and systemic mixing may alter the arterial supply more generally. Their coexistence does not determine their magnitudes. The anatomical and physiological questions have a substantial history \citep{Zenker1996,Caputa2004}, including explicit disagreement about selective brain cooling in humans \citep{White2011Point,Nybo2011Counterpoint}.

Controlled studies use different interventions and endpoints. Upper-airway airflow, intranasal balloon cooling and active evaporative devices have produced different local, cerebral and systemic temperature responses \citep{Andrews2005,Harris2007,Covaciu2011,AbouChebl2011,Nybo2014}. Their boundary conditions and measurement locations must be retained when relating those observations to a source-specific transfer coefficient.

The present study asks a restricted mathematical question. If a nasal-specific net heat export and a reduced allocation model are stipulated, which cerebral thermal-load contributions are compatible with the budget, and which combinations of coefficients can observations distinguish? This is a structural-identifiability question in the sense of uniqueness of parameters from ideal observations \citep{Bellman1970}. It precedes numerical fitting and statistical uncertainty estimation.

\subsection{Relation to existing thermal models}
Nasal computational fluid dynamics resolves airway conditioning, including spatial temperature, humidity and mucosal exchange \citep{Kim2017,Byun2019,Na2020}. Head bioheat models calculate tissue temperatures from perfusion, metabolism and thermal boundary conditions \citep{Sukstanskii2004,Sukstanskii2006}. \citet{Zhu2000} explicitly evaluated conduction and carotid--jugular countercurrent exchange in the neck, including upper thermal estimates. \citet{Lutz2019} coupled brain and systemic cooling during a specified intracarotid intervention. Consequently, combining heat-transfer pathways, using energy balances and calculating upper estimates are established approaches. Classical bioheat modeling \citep{Pennes1948} and the 83-node JOS-3 thermoregulation model \citep{Takahashi2021} also motivate explicit storage and individual-dependent inputs. The public JOS-3 documentation includes body dimensions, age and cardiac index among its inputs, and respiratory heat loss among its outputs \citep{JOS3Repository}. A more recent thermo-cardiovascular study reports individual temperature and circulatory predictions in a three-participant proof of concept \citep{Liu2026}. These forward-model variables do not independently identify a nasal-specific source or its cerebral contribution.

Table~\ref{tab:models} locates our question relative to these forward models. We study the attribution of a finite, specified nasal source under unknown effective couplings. The proposed results neither replace spatial transport calculations nor establish that the reduced allocation describes the human circulation. Their usefulness is to make the consequences of that allocation explicit, including the observations that can reject it or identify its effective coefficients.

\begin{table}[t]
\caption{Complementary thermal models and the source-attribution question. The comparison concerns the modeled input and output, not a claim that earlier models omit conservation or coupling.}
\label{tab:models}
\small
\begin{tabularx}{\linewidth}{@{}>{\raggedright\arraybackslash}p{0.20\linewidth}YY@{}}
\toprule
Study & Modeled input and output & Relation to nasal attribution\\
\midrule
\citet{Kim2017}; \citet{Byun2019} & Nasal geometry, airflow and wall conditions determine air conditioning and mucosal heat and water transfer. & Cerebral arterial thermal-load attribution is not the modeled output.\\[0.5em]
\citet{Sukstanskii2004,Sukstanskii2006} & Arterial temperature, perfusion, metabolism and environmental exchange determine head or brain temperature fields. & The origin of a prescribed arterial perturbation remains a separate input problem.\\[0.5em]
\citet{Zhu2000} & Blood and neck boundary temperatures, geometry and flow determine carotid cooling by conduction and countercurrent exchange. & Pathway contributions are evaluated without a measured nasal-specific net source budget.\\[0.5em]
\citet{Lutz2019} & An imposed intracarotid cooling intervention and perfusion conditions determine cerebral and systemic temperature time courses. & The arterial intervention is specified; coupling from an unknown nasal source is a different attribution problem.\\[0.5em]
\citet{Takahashi2021}; \citet{Liu2026} & Individual and environmental inputs determine whole-body thermal or circulatory responses. & Personalization does not supply an independently observed nasal-specific cerebral-power endpoint.\\[0.5em]
Present analysis & A specified nasal budget, effective couplings and optional residual-energy store determine output and inverse sets. & Static and transient results require matched inputs and independently reconstructed endpoints.\\
\bottomrule
\end{tabularx}
\end{table}

\subsection{Contributions and study design}
The analysis proceeds from steady allocation to controlled-condition identification and then to thermal storage. Section~\ref{sec:single} gives sharp bounds and exact single-endpoint inverse intervals. Section~\ref{sec:design} treats fixed flow fraction and bounded endpoint errors; Section~\ref{sec:variable-flow} allows known changes in that fraction. Section~\ref{sec:storage} proves integrated conservation and connects steady coupling with a transient time scale. The rank arguments are applications of linear algebra to the specified observation maps, not new general identifiability principles.

This is a theoretical modeling study using published aggregate evidence for context. It recruited no participants, collected no physiological data and fitted no human transfer coefficients. Synthetic examples test the algebra and computations. Exploratory scale calculations, an idealized conduction example and additional proofs are supplied separately. The effective models and their mathematical extensions require physiological validation before they can support human parameter estimates.

\section{Source, endpoint and allocation assumptions}
\label{sec:model}
\subsection{A nasal-specific net source}
Let $\dot q_{\mu\to a}(t)$ be the signed heat-transfer rate from nasal mucosa to respiratory air, including sensible and latent contributions. Over an observation window $T$ containing complete breaths, define
\begin{equation}
 N=\frac1T\int_0^T\dot q_{\mu\to a}(t)\,\dd t.
 \label{eq:source}
\end{equation}
The nonnegative-export regime $N\ge0$ is considered throughout. When the net transfer is into nasal tissue, the present nonnegative allocation requires reformulation. The algebraic allocation below applies to stationary conditions or observation windows for which storage changes are negligible. Section~\ref{sec:storage} treats finite storage explicitly.

To infer $N$ from humid-air enthalpy, measurements must resolve the nasal segment: appropriate upstream and downstream boundaries are needed in both flow directions, or an independently validated reconstruction of wall flux must be available. Ambient inspired air and terminal expired gas alone can include lower-airway exchange. Nor is inspiratory warming alone equal to net-cycle nasal export. These distinctions follow directly from the source definition and are consistent with the different experimental and numerical observables in \citep{Varene1986,Rouadi1999,Na2020}.

An external coolant, a unidirectional airflow apparatus and nasal inspiration followed by oral expiration change the boundary conditions. The symbol $N$ refers to one specified regime; it cannot be transferred unchanged between them. For an active device, a newly defined energy budget must include the device's heat extraction and any additional imposed sources or sinks.

\subsection{The cerebral endpoint}
We use $\Q$ for an \emph{attributed equivalent reduction in cerebral thermal load}, expressed as power. It is a quantity defined within the reduced allocation, not an absolute brain temperature, a whole-brain cooling rate obtained from one sensor, or a claim of selective cooling relative to core temperature. A physiological application must supply an independent reconstruction of an endpoint corresponding to this definition, with a common reference state and observation window.

For example, if cerebral mass flow and blood heat capacity are unchanged, an arterial inlet-temperature difference $\delta T_a$ corresponds to an inlet enthalpy perturbation
\begin{equation}
 \delta P_a=\dot m_b c_{p,b}\,\delta T_a.
 \label{eq:enthalpy-inlet}
\end{equation}
Equation~\eqref{eq:enthalpy-inlet} characterizes a change in incoming enthalpy relative to the stated reference. It does not identify a nasal cause or the net heat removed from the brain. A cerebral control-volume reconstruction must also account for venous outflow, storage, metabolism and nonvascular exchange. If flow changes, even the inlet term must be reconstructed from the corresponding enthalpy flows rather than a fixed-flow temperature conversion. In particular, multiplying a local subdural temperature change by total cerebral blood flow is not a valid endpoint reconstruction.

An arterial inlet measurement alone addresses the vascular contribution $\Q-C$ only when its enthalpy deficit can be attributed to the specified nasal source under the common reference conditions. Reconstructing the total endpoint $\Q$ also requires the independently determined conductive contribution $C$. If the vascular deficit itself is observed, it supplies $y_i=q_i-C_i$ directly in the repeated-condition equations below; $C_i$ must not be subtracted a second time.

The distinction between absolute attributed powers and intervention differences will be important in Section~\ref{sec:design}. A measurement of a difference between two conditions does not automatically supply the absolute $\Q$ in either condition.

\subsection{Exclusive allocation of the source}
All heat-rate symbols below have units of watts. Let $C\ge0$ be the part of the specified source attributed to direct conductive transfer from the cerebral region. Let $H\ge0$ be a stipulated regional enthalpy-deficit scale. A possible parameterization is
\begin{equation}
 H=\rho_b c_{p,b}\dot V_n\Delta T_v,
 \label{eq:H}
\end{equation}
where the flow, temperature difference and reference used for the regional stream must be defined independently. This parameterization does not assert that these quantities have been measured in a common human protocol.

Let $\alpha\in[0,1]$ describe the fraction of the regional scale routed through the proposed cerebral-relevant pathway and $\beta\in[0,1]$ its effective exchange fraction. Put $\gamma=\alpha\beta$. The regional allocation is $\gamma H$. We impose the shared-source constraint
\begin{equation}
 C+\gamma H\le N.
 \label{eq:budget}
\end{equation}
The residual $N-C-\gamma H$ is then available for the systemic allocation. Let $f$ be a specified cerebral fraction of cardiac output, and let $\eta\in[0,1]$ be an effective persistence fraction. The well-mixed allocation assumption assigns the fraction $s=f\eta$ of the residual to the cerebral endpoint. Under these steady-allocation assumptions,
\begin{equation}
 \boxed{\Q=C+\gamma H+s(N-C-\gamma H),\qquad s=f\eta.}
 \label{eq:model}
\end{equation}
The model avoids counting an allocated portion again in the residual. This is an accounting assumption; it is not a derivation of vascular routing, heat-exchanger efficiency or cardiovascular feedback from anatomy. Adiabatic mixing dilutes a temperature difference but does not destroy an enthalpy deficit. Accordingly, $\eta$ represents intervening heat exchange or compensatory rewarming relative to the specified reference, rather than an energy loss caused by mixing itself.

Table~\ref{tab:variables} records the domains. In the algebraic results, $H$ is a reference scale and may exceed $N-C$, provided the \emph{allocated} regional power obeys \eqref{eq:budget}. If instead $H$ is interpreted as the actual total nasal venous enthalpy deficit supplied solely by the same source, further local conservation restrictions may apply. For example, a closed steady allocation with only conductive and venous outflows would require $C+H\le N$. Such stronger restrictions must be imposed before physiological use; the weaker algebraic domain does not certify them.

\begin{table}[t]
\caption{Quantities in the reduced allocation. Knowing a domain does not supply an empirical value.}
\label{tab:variables}
\small
\begin{tabularx}{\linewidth}{@{}p{0.13\linewidth}Yp{0.22\linewidth}@{}}
\toprule
Symbol & Meaning within one specified regime & Domain\\
\midrule
$N$ & Net heat export attributable to nasal tissue & $N\ge0$\\
$C$ & Direct conductive allocation from the cerebral region & $0\le C\le N$\\
$H$ & Regional enthalpy-deficit reference scale & $H\ge0$\\
$\alpha,\beta$ & Effective routing and regional exchange fractions & $0\le\alpha,\beta\le1$\\
$\gamma$ & Product $\alpha\beta$, entering the endpoint & $0\le\gamma\le1$\\
$f,\eta$ & Cerebral flow fraction and systemic persistence fraction & $0\le f,\eta\le1$\\
$s$ & Effective systemic coefficient $f\eta$ & $0\le s\le f$\\
$B$ & Pre-systemic allocation $C+\gamma H$ & $C\le B\le N$\\
$\Q$ & Equivalent cerebral thermal-load allocation & Given by \eqref{eq:model}\\
\bottomrule
\end{tabularx}
\end{table}

Writing $B=C+\gamma H$ gives
\begin{equation}
 \Q=B+s(N-B)=sN+(1-s)B.
 \label{eq:convex}
\end{equation}
Consequently $B\le\Q\le N$ whenever the stated domain holds. The upper inequality is a property of this exclusive, nonnegative allocation. It is not a universal bound on all cerebral responses to a nasal intervention: additional heat sinks, altered metabolism, redistribution of other heat sources or regulatory changes fall outside the stipulated budget.

\section{Exact bounds and single-endpoint identification}
\label{sec:single}
\subsection{The full feasible output interval}
\begin{proposition}[Sharp conditional bounds]
\label{prop:bounds}
Fix $N\ge C\ge0$, $H\ge0$ and $0\le f\le1$. Under $0\le\gamma\le1$, $0\le s\le f$ and $C+\gamma H\le N$, put $D=\min\{H,N-C\}$. The set of values attained by \eqref{eq:model} is exactly
\begin{equation}
 \left[C,\ C+D+f(N-C-D)\right].
 \label{eq:sharp}
\end{equation}
\end{proposition}
\begin{proof}
Put $x=\gamma H$. The feasible values of $x$ form $[0,D]$: if $H>0$, every such value is realized by $\gamma=x/H$; if $H=0$, both $D$ and $x$ vanish. On the rectangle $[0,D]\times[0,f]$, the output is
\[
 F(x,s)=C+x+s(N-C-x).
\]
Its partial derivatives are $1-s\ge0$ and $N-C-x\ge0$. Its minimum is therefore $F(0,0)=C$, and its maximum is $F(D,f)=C+D+f(N-C-D)$. The rectangle is connected and $F$ is continuous, so its image contains every intermediate value.
\end{proof}

These are feasibility bounds conditional on the source scales, not confidence limits or population estimates. If $C$ is also unknown and its only restriction is $0\le C\le N$, the choice $C=\gamma=s=0$ is admissible. Thus the model and these domains alone do not establish a positive cerebral contribution.

\subsection{Complete conditional inverse intervals}
\begin{proposition}[Single-endpoint identification set]
\label{prop:single}
Let $N>C\ge0$, $H\ge0$, $0\le f<1$, $D=\min\{H,N-C\}$ and $U=C+D$. Let a proposed endpoint $q_*$ satisfy \eqref{eq:sharp}. The exact set of compatible systemic coefficients is $[s_-,s_+]$, where
\begin{equation}
 s_+=\min\left\{f,\frac{q_*-C}{N-C}\right\},\qquad
 s_-=\begin{cases}
 \displaystyle\max\left\{0,\frac{q_*-U}{N-U}\right\},&U<N,\\[0.6em]
 0,&U=N.
 \end{cases}
 \label{eq:single-interval}
\end{equation}
For every $s$ in this interval, the pre-systemic allocation is uniquely
\begin{equation}
 B(s)=\frac{q_*-sN}{1-s}.
 \label{eq:B-inverse}
\end{equation}
When $H>0$, its regional coefficient is $\gamma(s)=[B(s)-C]/H$; when $f>0$, its persistence fraction is $\eta=s/f$.
\end{proposition}
\begin{proof}
At fixed $s<1$, the map $B\mapsto sN+(1-s)B$ is strictly increasing. The condition $C\le B\le U$ is therefore equivalent to
\[
 C+s(N-C)\le q_*\le U+s(N-U).
\]
The left inequality gives the stated upper restriction on $s$. If $U<N$, the right inequality gives the stated lower restriction; if $U=N$, it reduces to $q_*\le N$. Intersecting with $[0,f]$ gives \eqref{eq:single-interval}. Conversely, each $s$ in that interval gives $C\le B(s)\le U$. If $H>0$, then $\gamma(s)\in[0,1]$ and $C+\gamma(s)H=B(s)\le N$. If $H=0$, feasibility forces $B(s)=C$ and the regional coefficient has no effect. This proves both necessity and sufficiency.
\end{proof}

\begin{corollary}[Regional attribution threshold]
\label{cor:regional}
Under Proposition~\ref{prop:single}, suppose $H>0$ and write $a=N-C$, $y=q_*-C$. The exact projection of the inverse set onto $\gamma$ is
\begin{equation}
 \max\left\{0,\frac{y-fa}{(1-f)H}\right\}
 \ \le\gamma\le\ \frac{\min\{D,y\}}{H}.
 \label{eq:gamma-interval}
\end{equation}
In particular, $q_*>C+f(N-C)$ forces $\gamma>0$ within the stipulated model.
\end{corollary}
\begin{proof}
For a fixed $x=\gamma H\in[0,D]$, varying $s\in[0,f]$ gives exactly
\[
 x\le y\le x+f(a-x)=fa+(1-f)x.
\]
Thus $\max\{0,(y-fa)/(1-f)\}\le x\le\min\{D,y\}$. Dividing by $H$ proves the interval, and every point is realizable by the preceding inequalities. If $x=a$, feasibility requires $y=a$ and any $s\in[0,f]$ realizes the endpoint; otherwise $s=(y-x)/(a-x)$ does so. The strict threshold follows immediately.
\end{proof}

The corollary specifies a useful distinction. A single endpoint generally fails to determine the effective pathway split, yet a sufficiently large endpoint could require a regional contribution if $N,C,H,f$ and the endpoint definition were independently established. Neither that threshold nor a numerical bound can be asserted from unmeasured source scales.

\subsection{Generic ambiguity and boundary cases}
For $N>C$, $H>0$ and $f>0$, the interval in \eqref{eq:single-interval} is generally nontrivial; the same endpoint then admits a continuum of systemic--regional allocations. This is not a universal impossibility of point identification. At the lowest endpoint $q_*=C$, for example, one must have $s=0$ and $\gamma=0$. If $f=0$, the regional allocation is $q_*-C$ and $\eta$ is unobservable. If $H=0$, the endpoint identifies $s=(q_*-C)/(N-C)$, while $\gamma$ is unobservable.

When $N=C$, feasibility requires $\gamma H=0$ and every admissible $s$ gives $\Q=C$. When $q_*=N$ and $f<1$, feasibility requires $B=N$; if $H>0$, this fixes $\gamma=(N-C)/H$ but leaves $s$ unrestricted. The algebraic saturation case $f=1$ is discussed in Supplementary Section~S1.

Even identified $\gamma$ generally fails to separate routing from exchange. For $0<\gamma<1$, all pairs
\begin{equation}
 \alpha\in[\gamma,1],\qquad \beta=\frac{\gamma}{\alpha}
 \label{eq:factorization}
\end{equation}
produce the same observations. At $\gamma=0$, either factor may vanish. At the exceptional boundary $\gamma=1$, the domain forces $\alpha=\beta=1$. Claims of non-identification of the separate factors must retain this exception.

\section{Identification across controlled conditions}
\label{sec:design}
\subsection{A linear observation map}
Consider $m$ conditions indexed by $i$. Suppose that $N_i,C_i,H_i$ and the corresponding absolute endpoint $q_i$ have been independently specified or reconstructed, and that the effective coefficients $\gamma$ and $s$ are constant over these conditions. Throughout this section, assume a common known $0<f<1$, so that $0\le s\le f$ and an identified $s$ also determines $\eta=s/f$. Constancy of the coefficients is an additional hypothesis, not a consequence of changing humidity or airflow.

Set
\begin{equation}
 a_i=N_i-C_i,\qquad y_i=q_i-C_i,\qquad t=(1-s)\gamma.
 \label{eq:linear-transform}
\end{equation}
Then
\begin{equation}
 y_i=s a_i+tH_i,\qquad
 X=\begin{pmatrix}a_1&H_1\\ \vdots&\vdots\\ a_m&H_m\end{pmatrix},\qquad
 \boldsymbol y=X\begin{pmatrix}s\\t\end{pmatrix}.
 \label{eq:linear-observations}
\end{equation}
The admissible coefficient domain is the closed convex polygon, possibly degenerate,
\begin{equation}
 \mathcal P=\left\{(s,t):
 0\le s\le f,\quad 0\le t\le1-s,\quad
 tH_i\le(1-s)a_i\ \text{for all }i\right\}.
 \label{eq:polygon}
\end{equation}
The last inequalities impose the source budget in every condition. Because $s\le f<1$, the transformation is invertible on its domain: $\gamma=t/(1-s)$.

\begin{proposition}[Finite-condition identification]
\label{prop:rank}
Under the stated constancy and known-input assumptions, the complete inverse set is
\[
 \left\{(s,t)\in\mathcal P:X(s,t)^{\mathsf T}=\boldsymbol y\right\}.
\]
If $\rank X=2$, this set contains at most one point. A point in the set identifies $s$, $\gamma=t/(1-s)$ and $\eta=s/f$. For two conditions, let
\[
 \Delta=a_1H_2-a_2H_1.
\]
When $\Delta\ne0$, the unique candidate is
\begin{equation}
 s=\frac{y_1H_2-y_2H_1}{\Delta},\qquad
 t=\frac{a_1y_2-a_2y_1}{\Delta}.
 \label{eq:inverse-two}
\end{equation}
If $\mathcal P$ has nonempty two-dimensional interior and a compatible coefficient lies in that interior, then $\rank X<2$ prevents its point identification.
\end{proposition}
\begin{proof}
The transformation \eqref{eq:linear-transform} changes the original domain exactly into \eqref{eq:polygon}, since $\gamma\le1$ becomes $t\le1-s$ and $\gamma H_i\le a_i$ becomes $tH_i\le(1-s)a_i$. It changes each endpoint equation exactly into \eqref{eq:linear-observations}. If $\rank X=2$, the kernel is zero, so two compatible coefficient vectors must coincide. Formula~\eqref{eq:inverse-two} follows by inverting the $2\times2$ matrix. Conversely, if $\rank X<2$, choose a nonzero $v\in\ker X$. At an interior compatible point $\theta$, the points $\theta+\varepsilon v$ remain in $\mathcal P$ for all sufficiently small $|\varepsilon|$ and yield the same data. Hence that point is not identified.
\end{proof}

The rank condition concerns distinguishability, not physiological plausibility. A rank-two candidate outside $\mathcal P$ is incompatible with the stipulated model. With more than two conditions, it must also reproduce all remaining endpoints. Conversely, rank deficiency can coexist with an isolated feasible point on a boundary; the proposition asserts necessity only for interior identification in a two-dimensional domain.

If $H_i=H>0$ is constant and $a_1\ne a_2$, the two rows are already independent. Thus independent changes in both inputs are not required. What matters is that the two columns of the actual design matrix are not proportional. Merely collecting more observations at a fixed ratio $H_i/a_i$ does not remove generic ambiguity.

\subsection{Conditioning and bounded endpoint errors}
The determinant in \eqref{eq:inverse-two} also exposes sensitivity. With the input matrix fixed and endpoint errors $|\delta y_i|\le\varepsilon_i$, direct subtraction gives
\begin{equation}
 |\delta s|\le\frac{|H_2|\varepsilon_1+|H_1|\varepsilon_2}{|\Delta|},\qquad
 |\delta t|\le\frac{|a_2|\varepsilon_1+|a_1|\varepsilon_2}{|\Delta|}.
 \label{eq:error-bounds}
\end{equation}
Near-proportional rows can therefore make a structurally identifiable design practically uninformative. For any two feasible candidates, $\gamma=t/(1-s)$ further satisfies
\begin{equation}
 |\gamma'-\gamma|\le
 \frac{|t'-t|+|s'-s|}{1-f}.
 \label{eq:gamma-error}
\end{equation}
Indeed, $\gamma'-\gamma=[t'-t+\gamma(s'-s)]/(1-s')$, with $0\le\gamma\le1$ and $1-s'\ge1-f$.

For observed endpoint intervals $\ell_i\le y_i\le u_i$, the exact coefficient set conditional on the input matrix is
\begin{equation}
 \mathcal F=\mathcal P\cap
 \bigcap_{i=1}^m\left\{(s,t):\ell_i\le a_i s+H_i t\le u_i\right\}.
 \label{eq:bands}
\end{equation}
It is a bounded intersection of linear half-planes. Its emptiness, vertices and projections can be computed without fitting a nonlinear physiological model. These are deterministic compatibility bands, not statistical confidence regions. Uncertainty in $N_i,C_i,H_i$ or $f$, and uncertainty in the validity of constant coupling, require an enlarged analysis; \eqref{eq:bands} does not account for them implicitly.

\subsection{A synthetic design example}
\begin{figure}[t]
\centering
\includegraphics[width=0.91\linewidth]{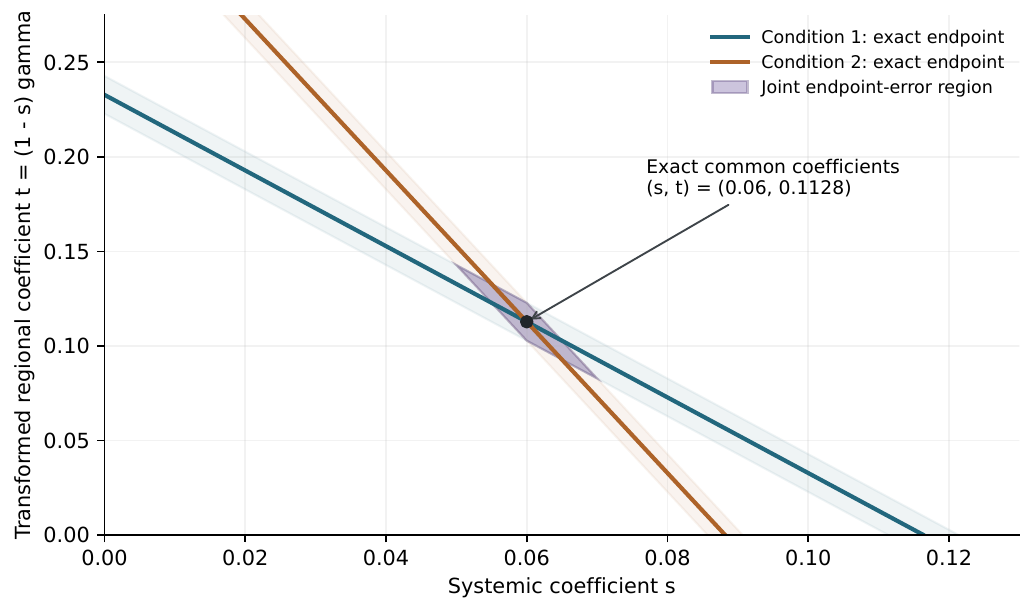}
\caption{Synthetic effective-coefficient identification. Each line represents one exact endpoint in Table~\ref{tab:synthetic}; their intersection identifies $(s,t)$. The shaded intersection uses deterministic $\pm0.01\W$ endpoint bands with the input matrix held fixed. The displayed window contains the entire endpoint-compatible region relevant to this example; the general prior domain is given by \eqref{eq:polygon}.}
\label{fig:design}
\end{figure}

Table~\ref{tab:synthetic} is an algebraic construction. Its inputs are chosen to illustrate identification and are not proposed human measurements. We use $f=0.15$, $s=0.06$ and $\gamma=0.12$, so $t=0.1128$ and $\eta=0.40$.
\begin{table}[ht]
\caption{Synthetic matched conditions. Heat rates are in watts solely to make the units explicit; these values have no empirical physiological status.}
\label{tab:synthetic}
\centering
\begin{tabular}{@{}lrrrrrr@{}}
\toprule
Condition & $N_i$ & $C_i$ & $H_i$ & $a_i$ & $y_i$ & $q_i$\\
\midrule
1 & 2.03 & 0.03 & 1.00 & 2.00 & 0.2328 & 0.2628\\
2 & 4.03 & 0.03 & 1.00 & 4.00 & 0.3528 & 0.3828\\
\bottomrule
\end{tabular}
\end{table}
Here $\Delta=-2\W^2$, and \eqref{eq:inverse-two} recovers the generating effective coefficients. Either observation alone admits multiple feasible pairs. Together they have a unique intersection (Fig.~\ref{fig:design}). Endpoint bands of $\pm0.01\W$ produce a small feasible polygon with $0.05\le s\le0.07$ and $0.0828\le t\le0.1428$. The two projection intervals do not describe independently selectable coefficients; their joint constraints must be retained. Even with exact data, the recovered $\gamma=0.12$ permits the continuum of routing--exchange factorizations in \eqref{eq:factorization}.

\subsection{What a perturbation slope identifies}
If $B=C+\gamma H$ and $s$ remain constant while $N$ varies, \eqref{eq:convex} becomes
\begin{equation}
 \Q(N)=sN+(1-s)B,\qquad \frac{\dd\Q}{\dd N}=s.
 \label{eq:slope}
\end{equation}
With two distinct values of $N$, known constant $C$ and $H>0$, absolute endpoints and the stated $s<1$, the slope identifies $s$, while the intercept identifies $B$ and hence $\gamma=(B-C)/H$. If only endpoint \emph{differences} are known, however, the constant offset $(1-s)B$ is not supplied by the data. In particular, a negligible response slope bounds $s$ under the invariance assumption; it does not bound the absolute regional or total contribution. The case $s=0$ still allows $\Q=B>0$.

If the coefficients or pre-systemic allocation vary with the perturbation, differentiation instead gives
\begin{equation}
 \frac{\dd\Q}{\dd N}
 =s+(1-s)\frac{\dd B}{\dd N}+(N-B)\frac{\dd s}{\dd N}.
 \label{eq:variable-slope}
\end{equation}
A small slope may then arise from compensating changes. Neither an approximately linear relationship nor a nonzero slope uniquely establishes a regional vascular mechanism.

\section{Identification with known variable cerebral flow fractions}
\label{sec:variable-flow}
The fixed-$f$ assumption can be relaxed when the cerebral flow fraction is known separately in each condition. Suppose $N_i\ge C_i\ge0$, $H_i\ge0$ and $0\le f_i\le1$ are known, and retain common $\eta,\gamma\in[0,1]$. Then
\begin{equation}
 q_i=C_i+\gamma H_i+f_i\eta(N_i-C_i-\gamma H_i),\qquad
 \gamma H_i\le N_i-C_i.
 \label{eq:variable-allocation}
\end{equation}
Here $s_i=f_i\eta$ can vary even when persistence and regional coupling remain constant. Physiological use requires matched cerebral and cardiac flows, with
\[
 f_i=\frac{\dot V_{\mathrm{cerebral},i}}{\dot V_{\mathrm{cardiac},i}},
 \qquad \dot V_{\mathrm{cardiac},i}>0.
\]
Body dimensions, age or a predicted core temperature do not independently determine these measured flows. Likewise, pooling individuals does not establish common $\eta$ and $\gamma$.

\subsection{A constrained linear representation}
Set $a_i=N_i-C_i$, $y_i=q_i-C_i$ and $\xi=\eta\gamma$. Define
\begin{equation}
 \boldsymbol\theta=\begin{pmatrix}\eta\\\gamma\\\xi\end{pmatrix},\qquad
 Z=\begin{pmatrix}
 f_1a_1&H_1&-f_1H_1\\
 \vdots&\vdots&\vdots\\
 f_ma_m&H_m&-f_mH_m
 \end{pmatrix}.
 \label{eq:lift}
\end{equation}
The linear equation $Z\boldsymbol\theta=\boldsymbol y$ must retain the product relation $\xi=\eta\gamma$.

\begin{proposition}[Exact inverse set for known variable flow]
\label{prop:variable-flow}
Under these assumptions, the complete coefficient set is the projection onto $(\eta,\gamma)$ of
\begin{equation}
 \mathcal L=\left\{(\eta,\gamma,\xi):
 \begin{array}{l}
 Z\boldsymbol\theta=\boldsymbol y,\quad 0\le\eta,\gamma\le1,\\
 \xi=\eta\gamma,\quad \gamma H_i\le a_i\ \text{for every }i
 \end{array}\right\}.
 \label{eq:lift-set}
\end{equation}
If $\rank Z=3$, there is at most one admissible pair. A unique linear candidate identifies the coefficients only if it satisfies all the constraints and all the observation equations.
\end{proposition}
\begin{proof}
Expanding \eqref{eq:variable-allocation} gives
$y_i=f_i a_i\eta+H_i\gamma-f_iH_i\eta\gamma$. Substitution of $\xi=\eta\gamma$ gives \eqref{eq:lift-set}. Conversely, every point of that set satisfies the original observations and domain, so the inverse sets agree exactly. If $Z$ has rank three, its kernel is zero; the difference of two compatible triples must therefore vanish.
\end{proof}

Rank three is sufficient, not necessary. When $f_i=f$, the third column of $Z$ is $-f$ times the second, so $\rank Z\le2$. Nevertheless, for $0<f<1$, the change of variables in Section~\ref{sec:design} can identify both coefficients. Identification on the product surface cannot be decided by requiring full rank in the three-dimensional representation.

\subsection{Two distinct flow fractions at fixed source scales}
\begin{proposition}[Two-condition flow design]
\label{prop:two-flow}
Suppose $a_1=a_2=a\ge0$, $H_1=H_2=H>0$ and $f_1\ne f_2$. Define
\begin{equation}
 d=\frac{y_2-y_1}{f_2-f_1},\qquad
 b=\frac{f_2y_1-f_1y_2}{f_2-f_1}.
 \label{eq:two-flow}
\end{equation}
The observations are compatible if and only if
\[
 0\le b\le\min(a,H),\qquad 0\le d\le a-b.
\]
For compatible observations, $\gamma=b/H$. If $a>b$, then $\eta=d/(a-b)$ is also identified. If $a=b$, compatibility requires $d=0$ and every $\eta\in[0,1]$ is admissible.
\end{proposition}
\begin{proof}
The two distinct abscissae $f_1,f_2$ determine the affine relation
$y_i=b+f_i d$. In the model its intercept and slope are
$b=\gamma H$ and $d=\eta(a-\gamma H)$. The domain and source budget imply the displayed bounds. Conversely, if those bounds hold and $a>b$, take $\gamma=b/H$ and $\eta=d/(a-b)$. These fractions satisfy the domain, budget and both equations. If $a=b$, the bounds force $d=0$; $\gamma=b/H$ and any $\eta$ then satisfy the equations. This also proves the identification statements.
\end{proof}

For example, the synthetic choice $a=2\W$, $H=1\W$, $\eta=0.4$, $\gamma=0.12$, $f_1=0.10$ and $f_2=0.20$ gives $y_1=0.1952\W$ and $y_2=0.2704\W$. The recovered intercept and slope are $b=0.12\W$ and $d=0.752\W$. This example specifies a distinguishable design, without asserting that a physiological intervention can vary $f$ while preserving the other quantities.

\subsection{Endpoint errors and degeneracies}
For fixed inputs and bands $\ell_i\le y_i\le u_i$, replace $Z\boldsymbol\theta=\boldsymbol y$ in \eqref{eq:lift-set} by the linear strips $\ell_i\le(Z\boldsymbol\theta)_i\le u_i$. The product constraint remains essential. The resulting set is generally nonconvex; deleting the product equation only gives an outer relaxation. Uncertain flows or source scales require additional variables and constraints.

If all $H_i=0$, the data contain no information about $\gamma$; if all $f_i=0$, they contain no information about $\eta$. Small $|f_2-f_1|$ amplifies endpoint errors in \eqref{eq:two-flow}. Identification of $\gamma$ still does not separate $\alpha$ and $\beta$, except at the boundary $\gamma=1$.

\section{Conservation with finite thermal storage}
\label{sec:storage}
The algebraic allocation describes a regime in which storage is negligible over the observation window. To examine what changes when a response is delayed, we introduce an effective residual-energy state. This is a specified extension of the accounting model; it is not a calibrated anatomical compartment.

Let $N(t),C(t),H(t)$ be nonnegative, locally integrable powers, and let $0\le\gamma(t),f(t)\le1$ be measurable, with
\begin{equation}
 B(t)=C(t)+\gamma(t)H(t)\le N(t),\qquad R(t)=N(t)-B(t)\ge0.
 \label{eq:residual-time}
\end{equation}
Here $N(t)$ is a time-resolved net export on a scale for which the nonnegative regime is justified, for example a sequence of complete-breath averages. It is not the generally signed instantaneous respiratory wall flux. Source boundaries and the reference state must remain consistent with Section~\ref{sec:model}.

Let $E(t)$ be residual source energy awaiting allocation, in joules, with $E(0)=E_0\ge0$. For constant transfer and rewarming rates $k_s,k_r\ge0$, in s$^{-1}$, assume $\sigma=k_s+k_r>0$ and set
\begin{align}
 E'(t)&=R(t)-\sigma E(t),\label{eq:reservoir}\\
 q_{\mathrm{dyn}}(t)&=B(t)+f(t)k_sE(t),\label{eq:dyn-output}\\
 O(t)&=(1-f(t))k_sE(t)+k_rE(t).\label{eq:other}
\end{align}
The quantity $O$ accounts for allocations outside the cerebral endpoint and for loss of the attributed thermal deficit through effective rewarming. The rates and the state interpretation are additional assumptions. No human rate is estimated here, and $E$ is not identified with an independently measured blood-pool energy deficit.

\begin{proposition}[Integrated conservation and stationary limit]
\label{prop:storage}
Under these assumptions, the unique absolutely continuous solution $E$ is nonnegative; $q_{\mathrm{dyn}}$ and $O$ are nonnegative almost everywhere. For every $T\ge0$,
\begin{equation}
 E(T)+\int_0^Tq_{\mathrm{dyn}}(t)\,\dd t+\int_0^TO(t)\,\dd t
 =E_0+\int_0^TN(t)\,\dd t.
 \label{eq:integrated}
\end{equation}
In particular,
\begin{equation}
 \int_0^Tq_{\mathrm{dyn}}(t)\,\dd t\le E_0+\int_0^TN(t)\,\dd t.
 \label{eq:integrated-bound}
\end{equation}
For constant $N,B,f$, the stationary output is
\begin{equation}
 q_{\mathrm{dyn},*}=B+f\eta(N-B),\qquad
 \eta=\frac{k_s}{k_s+k_r}.
 \label{eq:steady-recovery}
\end{equation}
Thus the original allocation is recovered at stationarity in this extension.
\end{proposition}
\begin{proof}
The integrating-factor formula gives the unique solution
\[
 E(t)=e^{-\sigma t}E_0+\int_0^t e^{-\sigma(t-r)}R(r)\,\dd r\ge0.
\]
The two outputs are nonnegative by their definitions. Adding \eqref{eq:reservoir}--\eqref{eq:other} gives $E'+q_{\mathrm{dyn}}+O=R+B=N$ almost everywhere. Integration proves \eqref{eq:integrated}; omitting the nonnegative storage and other-allocation terms proves \eqref{eq:integrated-bound}. Under constant forcing, $E_*=(N-B)/\sigma$ is the unique stationary state, approached exponentially. Substitution into \eqref{eq:dyn-output} proves \eqref{eq:steady-recovery}.
\end{proof}

\subsection{Why the instantaneous bound changes}
If the source is switched off at $t_0$ after positive energy has been stored, so that $N=B=0$ for $t>t_0$ and $E(t_0)>0$, then for constant $f>0$ and $k_s>0$,
\begin{equation}
 q_{\mathrm{dyn}}(t)=fk_sE(t_0)e^{-\sigma(t-t_0)}>0=N(t).
 \label{eq:tail}
\end{equation}
The output draws on an earlier input, consistently with \eqref{eq:integrated}. The static bound $\Q\le N$ therefore cannot be imposed pointwise on this transient model. For $T>0$, the corresponding bound on mean powers is $\overline q_T\le\overline N_T+E_0/T$; an unknown initial store contributes an unknown offset. If $E(T)=E_0$, the mean output over that window is at most the mean source. Signed responses, additional sinks or regulatory amplification require a different balance.

\begin{figure}[!b]
\centering
\includegraphics[width=0.88\linewidth]{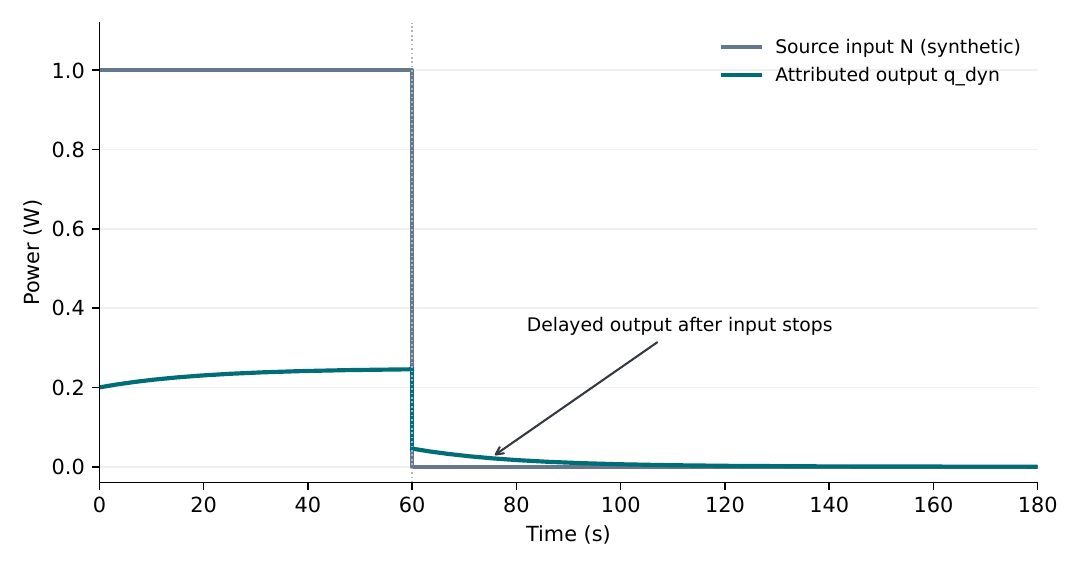}
\caption{Synthetic storage example. $N=1\W$ and $B=0.2\W$ for 60 s; both vanish thereafter. With $f=0.15$, $k_s=0.02$ s$^{-1}$, $k_r=0.03$ s$^{-1}$ and $E_0=0$, the attributed output persists after input cessation. All values are imposed for mathematical illustration, not physiological estimates.}
\label{fig:storage}
\end{figure}

The degenerate choices also have direct interpretations. If $k_s=0<k_r$, there is no delayed cerebral output and $\eta=0$; if $k_r=0<k_s$, then $\eta=1$. When both rates vanish, the separate system $E'=R$ and $q_{\mathrm{dyn}}=B$ still conserves energy, but the stationary quotient in \eqref{eq:steady-recovery} is undefined.

\subsection{Information supplied by a transient response}
Steady observations can determine the ratio $\eta=k_s/\sigma$ without determining the time scale $1/\sigma$. If $B(t)$, $N(t)$ and a positive $f(t)$ are independently known, a reconstructed output trajectory supplies
\[
 w(t)=\frac{q_{\mathrm{dyn}}(t)-B(t)}{f(t)}=k_sE(t),\qquad
 w'=k_sR-\sigma w.
\]
Integrating over observation intervals yields linear equations for $(k_s,\sigma)$. Two independent interval rows therefore determine at most one rate pair, subject to $0\le k_s\le\sigma$ and compatibility with the full trajectory. Supplementary Section~S10 gives the precise statement, proof and synthetic recovery example. This is conditional rate identification: an unknown $B$ or an unmeasured cerebral-power trajectory cannot be supplied by the same forward simulation and then treated as independent validation.

\section{Evidence and measurement requirements}
\label{sec:evidence}
\subsection{Observations from different regimes}
Table~\ref{tab:regimes} summarizes the distinctions needed to connect observations to the inverse problem. Positive cooling observations and limited or null selective effects are both relevant. Their interpretation depends on the location of the temperature measurement, the reference temperature, the intervention and the source boundary.

\begin{table}[t]
\caption{Experimental regimes and the information they supply. No row by itself provides the complete matched input--endpoint dataset required in Section~\ref{sec:design}.}
\label{tab:regimes}
\small
\begin{tabularx}{\linewidth}{@{}>{\raggedright\arraybackslash}p{0.22\linewidth}YY@{}}
\toprule
Regime and examples & Reported type of observation & Attribution limitation\\
\midrule
Respiratory conditioning \citep{Varene1986,Rouadi1999} & Expired-air heat and water exchange or imposed-flow nasopharyngeal conditioning. & Whole-respiratory loss or inspiratory conditioning does not isolate nasal-specific full-cycle $N$.\\[0.45em]
Local intracranial recordings \citep{Mariak1999} & Frontobasal temperature responses in postoperative patients, including nasal inspiration with oral expiration. & Local temperature does not supply whole-brain power or distinguish conduction from vascular transfer.\\[0.45em]
Upper-airway airflow in brain-injured patients \citep{Andrews2005,Harris2007} & Frontal parenchymal temperatures under different airflow protocols; limited or absent selective effects. & Protocols differ in exposure and co-interventions; a shared coupling coefficient cannot be assumed.\\[0.45em]
Active intranasal cooling in humans \citep{Covaciu2011,AbouChebl2011,Nybo2014} & Brain, core or arterial--jugular temperature responses, depending on the study. & Devices impose an external sink; endpoints and selectivity criteria differ from spontaneous breathing.\\[0.45em]
High-flow or device-based animal interventions \citep{Stonko2023,Koehler2023,Nishikimi2026} & Cerebral cooling under forced airflow, active cooling or combined cooling and systemic warming. & Anatomy, imposed flows and additional sources or sinks limit transfer to the passive human budget.\\
\bottomrule
\end{tabularx}
\end{table}

\citet{Mariak1999} provide direct local intracranial observations, but their respiratory maneuver and measurement sites differ from both ordinary bidirectional breathing and the trials of \citet{Andrews2005,Harris2007}. A conduction-only half-space example in Supplementary Section~S5 can produce a delayed local temperature pulse within the reported aggregate ranges using prescribed, unmeasured boundary and distance parameters. Its role is a mathematical example of non-unique attribution from limited summary observations. It is not a fit to patient trajectories, an anatomical model or a demonstration that conduction caused the observed response.

Device studies likewise demonstrate that cooling can occur under specified interventions, without identifying the contribution of ordinary nasal heat exchange. The MR endpoints of \citet{Covaciu2011}, the arterial--jugular measurements of \citet{Nybo2014} and the clinical feasibility observations of \citet{AbouChebl2011} must retain their different definitions. The animal studies in Table~\ref{tab:regimes} add evidence under further boundary conditions. In particular, the simultaneous systemic warming in the protocol of \citet{Nishikimi2026} is part of that intervention and cannot be removed from its interpretation.

\subsection{A staged identification protocol}
The algebra suggests a sequence of measurement tasks rather than a claim that all required quantities are presently accessible.

\paragraph{Define and reconstruct the source.}
Specify one respiratory route and observation window. Measure heat and water exchange at boundaries that separate the nasal segment, or validate an independent nasal wall-flux reconstruction. Integrate over complete cycles and account for storage if the observation is transient. Keep device heat extraction separate when an active intervention is used.

\paragraph{Establish the endpoint and reference.}
Reconstruct a power endpoint that matches the observation map: either the total attributed contribution $q_i$, including conduction, or the vascular contribution $y_i=q_i-C_i$. An arterial inlet-temperature change alone does not supply the total endpoint. Record the reference state and relevant flow changes. Temperature alone is insufficient, and the measurements used to define the source must not simply be recycled as a model-generated endpoint. If only relative effects can be reconstructed, formulate the inverse problem for those differences and retain the unresolved offset.

\paragraph{Constrain the auxiliary scales.}
Obtain independent information on $C_i$ and $H_i$, with their uncertainty and local energy interpretation. Regional flow and venous temperature must be matched to the same condition. A local perfusion value per unit mucosal mass does not measure total regional flow; the geometric and perfusion extrapolations discussed in Supplementary Sections~S2--S3 do not provide that calibration. Measure cerebral and cardiac flow in each condition if $f$ or $\eta$ is to be interpreted physiologically. Whole-body forward-model outputs may help formulate a reconstruction, but they are not independent observations; total respiratory heat loss must not be substituted for the nasal-specific source.

\paragraph{Design distinguishable conditions.}
A controlled humidity change at approximately fixed inspired temperature and ventilation is one possible way to change nasal heat export. Whether it preserves $s$ and $\gamma$ must be tested or independently constrained, because perfusion and vascular tone may also change. Before interpreting fitted coefficients, examine $\rank X$ and its conditioning. Known $H>0$ with distinct $N_i-C_i$ is sufficient algebraically. If $f_i$ changes and is measured, use the observation map in Section~\ref{sec:variable-flow} and assess the remaining constancy assumptions. Neither design establishes that a proposed intervention realizes the required invariances.

\paragraph{Use additional conditions to test compatibility.}
For the fixed-flow design, two independent conditions can determine a candidate. The two-flow construction has its own fixed-source requirements. Additional conditions allow prediction checks with the same coefficients. Incompatible exact data, or an empty bounded-error set \eqref{eq:bands}, reject the combined source, endpoint and constant-coupling assumptions at the stated tolerances. They do not by themselves identify which assumption failed. Compatible data constrain the model but do not prove an anatomical route. Separating $\alpha$ and $\beta$ away from the boundary requires information beyond their product. For transient observations, the initial store and time-resolved inputs must also be constrained; an output after input cessation is compatible with storage and does not alone identify its physiological mechanism.

\section{Discussion}
The framework distinguishes energetic compatibility, effective-coefficient identification and physiological source attribution. A nonnegative allocation can satisfy its source budget while admitting several pathway decompositions. A single endpoint can sometimes require a positive regional contribution without uniquely identifying its magnitude. Even a unique effective regional product generally leaves routing and exchange inseparable. These conclusions follow from the specified observation map and its domain; they do not depend on assigning plausible-looking values to unmeasured fractions.

Known variation in cerebral flow fraction changes the observation map and can add information. It also changes which coefficients are assumed invariant. A fixed $s=f\eta$ design and a fixed $\eta$ design with varying $f$ are different experiments. Individualized thermophysiological models can help represent background states, but their demographic inputs or predicted temperatures do not establish those invariances. In particular, pooling participants under shared $\eta$ and $\gamma$ requires independent support. The product constraint in the variable-flow inverse set must be retained when evaluating fit or uncertainty.

Storage introduces a second distinction. Steady data can constrain allocation fractions without identifying a transport time scale; transient data can provide that additional information under known inputs. Conservation then constrains integrated energy, including the initial store, rather than necessarily bounding output by the simultaneous source. The one-store construction demonstrates this difference with an explicit balance and stationary limit. It does not identify a blood compartment, validate exponential decay in humans or rule out other delay mechanisms. The conduction example in the supplement illustrates a distinct mechanism capable of producing a delayed local temperature response.

The principal limitation is the need for independently constrained inputs and a meaningful cerebral-power endpoint. $N$ must isolate the nasal source, $C$ must represent a separately reconstructed conductive contribution, and $H$ must have a defined stream and reference interpretation. These requirements are mathematical conditioning assumptions, not evidence that all quantities can presently be measured accurately in one protocol. Source uncertainty, unknown initial storage and uncertain flow fractions enlarge the inverse problem. The endpoint-only polygon in Section~\ref{sec:design} must not be interpreted as accounting for these additional uncertainties.

The model also excludes signed pathway effects, regulatory amplification and detailed spatial transport. Nasal CFD, head bioheat models and systemic thermal networks provide complementary ways to examine those processes \citep{Kim2017,Byun2019,Sukstanskii2004,Zhu2000,Lutz2019,Takahashi2021}. Connecting a network to this allocation requires an explicit source partition and endpoint map; Supplementary Section~S9 gives the relevant conservation requirements. Additional transport detail may impose stronger feasibility constraints or invalidate the reduced observation equation.

The supplied computations distinguish algebraic verification from empirical validation. Synthetic designs test exact recovery, bounded-error compatibility and conservation under the imposed equations. The exploratory physiological-scale calculations retained in the supplement remain conditional constructions. Neither their numerical size nor agreement with a few published local-temperature summaries estimates natural human cooling capacity. The practical contribution is a set of explicit checks and measurement requirements that can be applied before interpreting a fitted thermal response as evidence for a nasal-to-cerebral pathway.

\section{Conclusion}
A specified nasal heat source admits sharp conditional output bounds and exact inverse sets under exclusive conductive, regional and systemic allocation. Repeated matched conditions identify effective coefficients when the observation map is sufficiently informative and its invariance assumptions hold. Known variation in cerebral flow fraction permits an additional design, with an essential product constraint. Introducing finite residual storage replaces the pointwise source bound by an integrated energy balance and recovers the steady model at stationarity. Time-resolved data can further distinguish a rate ratio from a transport time scale.

Physiological use requires a common protocol that establishes the nasal source, auxiliary scales, flow fractions and cerebral-power endpoint independently. Published temperatures from different interventions cannot supply those quantities by substitution. The analysis specifies the inferences supported by the stated model and the observations needed to test it; a positive natural human cooling rate and therapeutic efficacy remain unestablished by this study.

\section*{Statements and Declarations}
\paragraph{Funding.} This research received no external funding.
\paragraph{Competing interests.} The authors declare no competing financial or non-financial interests.
\paragraph{Ethics and consent.} This theoretical study used published aggregate results. It involved no recruitment, interventions, new clinical observations, participant-level datasets or clinical photographs.
\paragraph{Data and code availability.} The accompanying source archive contains the LaTeX manuscript and supplement, synthetic inputs, numerical reproduction scripts, and the data used to generate the figures. No individual participant data were obtained.
\ifshowauthors
\paragraph{Author contributions.} William \'{A}lvarez Gaviria: conceptualization, physiological hypothesis and anatomical interpretation. Juan D. V\'{e}lez: methodology, mathematical modeling and formal analysis. Both authors contributed to manuscript development.
\else
\paragraph{Author contributions.} Author information is supplied in the separate title page.
\fi
\paragraph{AI-assisted preparation.} OpenAI ChatGPT was used to assist with literature discovery, language editing, manuscript organization, LaTeX preparation, algebraic derivations and numerical verification scripts. Synthetic computations are disclosed as such and the accompanying code makes them reproducible. The authors are responsible for checking the final text, references and analyses and for the content of the submitted version.


\begin{thebibliography}{99}
\small
\bibitem[Abou-Chebl et al.(2011)]{AbouChebl2011}Abou-Chebl A, Sung G, Barbut D, Torbey M. Local brain temperature reduction through intranasal cooling with the RhinoChill device: preliminary safety data in brain-injured patients. \textit{Stroke}. 2011;42(8):2164--2169. \url{https://doi.org/10.1161/STROKEAHA.110.613000}.
\bibitem[Andrews et al.(2005)]{Andrews2005}Andrews PJD, Harris B, Murray GD. Randomized controlled trial of effects of the airflow through the upper respiratory tract of intubated brain-injured patients on brain temperature and selective brain cooling. \textit{Br J Anaesth}. 2005;94:330--335. \url{https://doi.org/10.1093/bja/aei025}.
\bibitem[Bellman and \AA{}str\"om(1970)]{Bellman1970}Bellman R, \AA{}str\"om KJ. On structural identifiability. \textit{Math Biosci}. 1970;7(3--4):329--339. \url{https://doi.org/10.1016/0025-5564(70)90132-X}.
\bibitem[Byun et al.(2019)]{Byun2019}Byun S, Chung S-K, Na Y. Air-conditioning characteristics in nasal cavity models exhibiting nasal cycle states. \textit{J Therm Biol}. 2019;83:60--68. \url{https://doi.org/10.1016/j.jtherbio.2019.05.004}.
\bibitem[Caputa(2004)]{Caputa2004}Caputa M. Selective brain cooling: a multiple regulatory mechanism. \textit{J Therm Biol}. 2004;29(7--8):691--702. \url{https://doi.org/10.1016/j.jtherbio.2004.08.079}.
\bibitem[Covaciu et al.(2011)]{Covaciu2011}Covaciu L, Weis J, Bengtsson C, Allers M, Lunderquist A, Ahlstr\"om H, Rubertsson S. Brain temperature in volunteers subjected to intranasal cooling. \textit{Intensive Care Med}. 2011;37(8):1277--1284. \url{https://doi.org/10.1007/s00134-011-2264-7}.
\bibitem[Harris et al.(2007)]{Harris2007}Harris BA, Andrews PJD, Murray GD. Enhanced upper respiratory tract airflow and head fanning reduce brain temperature in brain-injured, mechanically ventilated patients: a randomized, crossover, factorial trial. \textit{Br J Anaesth}. 2007;98(1):93--99. \url{https://doi.org/10.1093/bja/ael308}.
\bibitem[Kim et al.(2017)]{Kim2017}Kim D-W, Chung S-K, Na Y. Numerical study on the air conditioning characteristics of the human nasal cavity. \textit{Comput Biol Med}. 2017;86:18--30. \url{https://doi.org/10.1016/j.compbiomed.2017.04.018}.
\bibitem[Koehler et al.(2023)]{Koehler2023}Koehler RC, Reyes M, Hopkins CD, et al. Rapid, selective and homogeneous brain cooling with transnasal flow of ambient air for pediatric resuscitation. \textit{J Cereb Blood Flow Metab}. 2023;43(11):1842--1856. \url{https://doi.org/10.1177/0271678X231189463}.
\bibitem[Liu et al.(2026)]{Liu2026}Liu H, Ogura D, Hokoi S, Iba C. An improved thermo-cardiovascular model for individual thermophysiological prediction in non-uniform indoor thermal environments. \textit{Build Simul}. 2026;19(3):809--832. \url{https://doi.org/10.1007/s12273-026-1419-y}.
\bibitem[Lutz et al.(2019)]{Lutz2019}Lutz Y, Loewe A, Meckel S, D\"ossel O, Cattaneo G. Combined local hypothermia and recanalization therapy for acute ischemic stroke: estimation of brain and systemic temperature using an energetic numerical model. \textit{J Therm Biol}. 2019;84:316--322. \url{https://doi.org/10.1016/j.jtherbio.2019.06.011}.
\bibitem[Mariak et al.(1999)]{Mariak1999}Mariak Z, White MD, Lewko J, Lyson T, Piekarski P. Direct cooling of the human brain by heat loss from the upper respiratory tract. \textit{J Appl Physiol}. 1999;87:1609--1613. \url{https://doi.org/10.1152/jappl.1999.87.5.1609}.
\bibitem[Na et al.(2020)]{Na2020}Na Y, Chung S-K, Byun S. Numerical study on the heat-recovery capacity of the human nasal cavity during expiration. \textit{Comput Biol Med}. 2020;126:103992. \url{https://doi.org/10.1016/j.compbiomed.2020.103992}.
\bibitem[Nishikimi et al.(2026)]{Nishikimi2026}Nishikimi M, Kikutani K, Higashi M, Ohshimo S, Anzai T, Shime N. Exploratory study of selective brain hypothermia using transnasal evaporative cooling under controlled normothermia with an endovascular device. \textit{J Cardiovasc Dev Dis}. 2026;13(3):120. \url{https://doi.org/10.3390/jcdd13030120}.
\bibitem[Nybo and Secher(2011)]{Nybo2011Counterpoint}Nybo L, Secher NH. Counterpoint: Humans do not demonstrate selective brain cooling during hyperthermia. \textit{J Appl Physiol}. 2011;110:571--573. \url{https://doi.org/10.1152/japplphysiol.00992.2010a}.
\bibitem[Nybo et al.(2002)]{Nybo2002}Nybo L, Secher NH, Nielsen B. Inadequate heat release from the human brain during prolonged exercise with hyperthermia. \textit{J Physiol}. 2002;545(2):697--704. \url{https://doi.org/10.1113/jphysiol.2002.030023}.
\bibitem[Nybo et al.(2014)]{Nybo2014}Nybo L, Wanscher M, Secher NH. Influence of intranasal and carotid cooling on cerebral temperature balance and oxygenation. \textit{Front Physiol}. 2014;5:79. \url{https://doi.org/10.3389/fphys.2014.00079}.
\bibitem[Pennes(1948)]{Pennes1948}Pennes HH. Analysis of tissue and arterial blood temperatures in the resting human forearm. \textit{J Appl Physiol}. 1948;1(2):93--122. \url{https://doi.org/10.1152/jappl.1948.1.2.93}.
\bibitem[Rouadi et al.(1999)]{Rouadi1999}Rouadi P et al. A technique to measure the ability of the human nose to warm and humidify air. \textit{J Appl Physiol}. 1999;87:400--406. \url{https://doi.org/10.1152/jappl.1999.87.1.400}.
\bibitem[Stonko et al.(2023)]{Stonko2023}Stonko DP et al. High flow cooled air can decrease brain temperature without injuring the snout or brain in swine. \textit{Surg Pract Sci}. 2023;12:100154. \url{https://doi.org/10.1016/j.sipas.2022.100154}.
\bibitem[Sukstanskii and Yablonskiy(2004)]{Sukstanskii2004}Sukstanskii AL, Yablonskiy DA. An analytical model of temperature regulation in human head. \textit{J Therm Biol}. 2004;29(7--8):583--587. \url{https://doi.org/10.1016/j.jtherbio.2004.08.028}.
\bibitem[Sukstanskii and Yablonskiy(2006)]{Sukstanskii2006}Sukstanskii AL, Yablonskiy DA. Theoretical model of temperature regulation in the brain during changes in functional activity. \textit{Proc Natl Acad Sci USA}. 2006;103(32):12144--12149. \url{https://doi.org/10.1073/pnas.0604376103}.
\bibitem[Takahashi et al.(2021)]{Takahashi2021}Takahashi Y, Nomoto A, Yoda S, Hisayama R, Ogata M, Ozeki Y, Tanabe S. Thermoregulation model JOS-3 with new open source code. \textit{Energy Build}. 2021;231:110575. \url{https://doi.org/10.1016/j.enbuild.2020.110575}.
\bibitem[Tanabe Laboratory(n.d.)]{JOS3Repository}Tanabe Laboratory. JOS-3: official source repository and documentation. \url{https://github.com/TanabeLab/JOS-3}. Accessed 24 November 2026. The documentation was consulted; the software was not executed in this study.
\bibitem[Var\`ene et al.(1986)]{Varene1986}Var\`ene P, Ferrus L, Manier G, Gire J. Heat and water respiratory exchanges: comparison between mouth and nose breathing in humans. \textit{Clin Physiol}. 1986;6:405--414. \url{https://doi.org/10.1111/j.1475-097X.1986.tb00071.x}.
\bibitem[White et al.(2011)]{White2011Point}White MD, Greiner JG, McDonald PLL. Point: Humans do demonstrate selective brain cooling during hyperthermia. \textit{J Appl Physiol}. 2011;110:569--571. \url{https://doi.org/10.1152/japplphysiol.00992.2010}.
\bibitem[Zenker and Kubik(1996)]{Zenker1996}Zenker W, Kubik S. Brain cooling in humans---anatomical considerations. \textit{Anat Embryol}. 1996;193(1):1--13. \url{https://doi.org/10.1007/BF00186829}.
\bibitem[Zhu(2000)]{Zhu2000}Zhu L. Theoretical evaluation of contributions of heat conduction and countercurrent heat exchange in selective brain cooling in humans. \textit{Ann Biomed Eng}. 2000;28:269--277. \url{https://doi.org/10.1114/1.266}.
\end{thebibliography}
\end{document}

\maketitle
\noindent This supplement contains algebraic edge cases, exploratory scale calculations, an idealized conduction construction, a compartmental interpretation, transient rate-identification proofs and reproducibility details. None of the numerical scenarios calibrates a human nasal-to-cerebral transfer coefficient. The main manuscript contains the source and endpoint definitions and the principal steady, variable-flow and energy-storage results.

\section{Degenerate cases of the inverse problem}
For reference, the allocation is
\begin{equation}
 \Q=C+\gamma H+s(N-C-\gamma H),\qquad
 0\le s\le f,\quad 0\le\gamma\le1,\quad C+\gamma H\le N.
 \label{eq:S-model}
\end{equation}
Here $s=f\eta$ and $\gamma=\alpha\beta$, with all original fractions in $[0,1]$. Let $a=N-C$, $D=\min(H,a)$ and $U=C+D$.

\begin{table}[ht]
\caption{Degenerate or boundary observations. Each row is conditional on all other stipulated inputs being known and on compatibility with the model.}
\small
\begin{tabularx}{\linewidth}{@{}>{\raggedright\arraybackslash}p{0.29\linewidth}Y@{}}
\toprule
Condition & Consequence\\
\midrule
$N=C$ & $\gamma H=0$ and $\Q=C$; $s$ is unrestricted. If $H>0$, $\gamma=0$.\\[0.4em]
$H=0$, $N>C$ & $s=(\Q-C)/(N-C)$; $\gamma$ does not affect the endpoint.\\[0.4em]
$f=0$, $H>0$ & $s=0$ and $\gamma=(\Q-C)/H$; $\eta$ is unrestricted.\\[0.4em]
$\Q=C$, $N>C$, $H>0$ & $s=0$ and $\gamma=0$, despite having only one endpoint.\\[0.4em]
$U<N$, $H>0$, $\Q=U+f(N-U)$, $f<1$ & $s=f$ and $\gamma=1$. Here $U=C+H$.\\[0.4em]
$U=N$, $N>C$, $H>0$, $\Q=N$, $f<1$ & $\gamma=(N-C)/H$ and any $s\in[0,f]$ is admissible.\\[0.4em]
$\gamma=1$ & $\alpha=\beta=1$. For $0<\gamma<1$, factorization is nonunique; for $\gamma=0$, at least one factor vanishes.\\
\bottomrule
\end{tabularx}
\end{table}

The main single-endpoint formula assumes $f<1$ so that $1-s$ never vanishes. If $f=1$ and $\Q<N$, one still has $s<1$; the same formula applies with $s_+=\min\{1,(\Q-C)/(N-C)\}<1$ when $N>C$. At the saturation endpoint $\Q=N$, the condition becomes
\[
 (1-s)(N-B)=0,\qquad B=C+\gamma H.
\]
The compatible set always contains $s=1$ with any $B\in[C,U]$. If and only if $U=N$, it also contains $B=N$ with any $s\in[0,1]$. The inverse formula dividing by $1-s$ must not be used on the first branch.

Boundary point identification also occurs for a rank-deficient repeated-condition design. With rows $(a_1,H_1)=(2,1)$ and $(a_2,H_2)=(4,2)$, the matrix has rank one. Yet endpoints $y_1=y_2=0$ force $s=t=0$ because $s,t\ge0$. This is why the necessity statement for rank two in the main text is restricted to interior points of a two-dimensional feasible domain.

\section{Nasal heat-processing scales and input provenance}
\subsection{An exploratory sensible--latent balance}
A simplified full-cycle heat-export calculation, after nasal attribution and expiratory recovery have been established, has the form
\begin{equation}
 N=\rho_a\dot V_a c_{p,a}\Delta T_a+L_v\dot m_w.
 \label{eq:S-source}
\end{equation}
Here $\Delta T_a$ and $\dot m_w$ must represent the appropriate net nasal contributions, not inspiratory conditioning alone. Temperature-dependent humid-air enthalpy corrections and any additional tissue storage must be included when required by the protocol. Whole-respiratory measurements \citep{Varene1986} and theoretical water-loss estimates \citep{Zielinski2012} do not by themselves supply this nasal-specific source.

An illustrative calculation uses $\dot V_a=8$ L/min, $\rho_a=1.18$ kg/m$^3$, $c_{p,a}=1005$ J/(kg K), $\Delta T_a=8$ K, $\dot m_w=0.09$ g/min and $L_v=2.50\times10^6$ J/kg. Unit conversion gives
\begin{align}
 \dot V_a&=1.33333\times10^{-4}\ \mathrm{m^3/s},&
 \dot m_w&=1.50\times10^{-6}\ \mathrm{kg/s},\\
 \Q_{\mathrm{sens}}&=1.26496\W,&
 \Q_{\mathrm{lat}}&=3.75\W.
 \label{eq:S-powers}
\end{align}
Their unrounded sum is $5.01496\W$. For reproducibility, the reference scenario uses the separately rounded working value $N=5.02\W$, which is retained explicitly below to reproduce that scenario. This minor rounding distinction has no physiological significance; neither value is a measured nasal export.

The humidity-ratio representation is
\begin{equation}
 \omega=0.622\frac{p_v}{P-p_v},\qquad p_v=\phi p_{\mathrm{sat}}(T),\qquad
 \Q_{\mathrm{lat}}=L_v\dot m_{\mathrm{da}}\Delta\omega.
 \label{eq:S-humidity}
\end{equation}
It expresses the dependence on temperature, humidity and dry-air mass flow. A numerical prediction still requires the relevant local conditions and the full-cycle nasal attribution. Neither a simulated recovery ratio \citep{Na2020} nor inspiratory air-conditioning efficiency \citep{Naftali2005} is a universal correction for an independently assumed source.

\subsection{Geometric scales do not identify a transfer coefficient}
An approximate nasal surface area of $160$ cm$^2$ \citep{Gizurarson2012} and an air-filled volume near $17$ mL, consistent with the CT scales in \citep{Loftus2016}, give an area-to-volume ratio near $9.4$ cm$^{-1}$. Dividing $17$ mL by an assumed $8$ L/min gives a kinematic renewal scale of about $0.128$ s. This quotient is neither an exact residence time in a pulsatile complex airway nor evidence of cerebral heat-transfer efficiency. Air-filled volume also does not determine perfused mucosal mass.

Table~\ref{tab:S-provenance} separates measured quantities in particular studies from the assumptions needed to construct the example. A measurement from one anatomical site or regime is not a measured coefficient for the entire transfer chain.

\begin{table}[p]
\caption{Provenance of exploratory inputs. The entries describe evidential roles, not statistical distributions.}
\label{tab:S-provenance}
\small
\begin{tabularx}{\linewidth}{@{}>{\raggedright\arraybackslash}p{0.27\linewidth}>{\raggedright\arraybackslash}p{0.22\linewidth}Y@{}}
\toprule
Quantity & Working value & Status and limitation\\
\midrule
Nasal surface area & $160$ cm$^2$ & Anatomical scale \citep{Gizurarson2012}; not a measured active cerebral exchange area.\\[0.5em]
Respiratory epithelial area & $120$ cm$^2$ & Anatomical description \citep{Beule2010}; distinct from the broader surface estimate.\\[0.5em]
Air-filled nasal volume & $17$ mL & CT-derived scale \citep{Loftus2016}; not perfused tissue volume.\\[0.5em]
Ventilation & $8$ L/min & Assumed reference condition.\\[0.5em]
Net nasal sensible difference & $8$ K & Assumed full-cycle nasal contribution; not measured here.\\[0.5em]
Net nasal water export & $0.09$ g/min & Assumed nasal attribution; whole-respiratory calculations do not identify it \citep{Zielinski2012}.\\[0.5em]
Mass-specific mucosal perfusion & $33\pm10$ mL/min per $100$ g & Reported mean and between-subject SD in $^{133}$Xe measurements \citep{Bende1983}; not total nasal flow.\\[0.5em]
Mucosal thickness & $0.3$--$5$ mm & Heterogeneous anatomical range \citep{Beule2010}; its use as one uniform effective depth is an extrapolation.\\[0.5em]
Hyperthermic flow multiplier & $3$ & Approximate local relative response \citep{White1995}; uniform whole-nose scaling is unverified.\\[0.5em]
Regional flow and venous decrement & $50$ mL/min; $1$ K & Unmeasured working inputs to $H$.\\[0.5em]
Conductive allocation & $C=0.05\W$ & Unmeasured illustrative value; not inferred from the transient construction in S5.\\[0.5em]
Routing and exchange fractions & $\alpha=\beta=0.20$ & Unmeasured assumptions; $\gamma=0.04$.\\[0.5em]
Systemic persistence & $\eta=0.45$ & Unmeasured assumption.\\[0.5em]
Cerebral flow share & $f=0.15$ & Assumed physiological scale; no matched flow measurement.\\
\bottomrule
\end{tabularx}
\end{table}

\section{Exploratory vascular and coupling calculations}
\subsection{A conditional perfusion--geometry construction}
\citet{Bende1983} reported mass-specific nasal mucosal perfusion of $33\pm10$ mL/min per $100$ g (mean $\pm$ SD) in 80 participants. \citet{White1995} observed an approximately threefold local relative laser-Doppler response during passive hyperthermia in five subjects. These methods and observables differ. Combining their scales over an assumed uniform tissue mass is a constructed extrapolation.

Write $M_R=\rho_t A_Rd_R$, with area in cm$^2$, depth in cm and density in g/cm$^3$. Applying the mean perfusion $0.33$ mL/min per g uniformly, with a further uniform multiplier $r$, gives
\begin{equation}
 \dot V_n=0.33rM_R\ \mathrm{mL/min},\qquad
 H=0.020988\,rM_R\Delta T_v\ \mathrm{W},
 \label{eq:S-mass}
\end{equation}
where the latter expression uses $\rho_b=1060$ kg/m$^3$, $c_{p,b}=3600$ J/(kg K), $M_R$ in g and $\Delta T_v$ in K. Neither the regional routing fraction nor the actual thermal deficit is identified by this construction.

For $A_R=120$ cm$^2$ and $\rho_t=1$ g/cm$^3$, treating a heterogeneous $0.3$--$5$ mm thickness range as a uniform effective depth gives $3.6\le M_R\le60$ g. The resulting flow ranges are $1.188$--$19.8$ mL/min for $r=1$ and $3.564$--$59.4$ mL/min for $r=3$. These are algebraic outputs of the extrapolation, not observed ranges or population confidence intervals. Site-specific turbinate measurements \citep{Berger2000} likewise cannot be substituted for a whole-nose effective depth.

The reference $50$ mL/min would require $M_R\simeq50.5$ g and $d_R\simeq4.21$ mm under $r=3$ and the preceding uniform assumptions. Table~\ref{tab:S-mass} shows the corresponding scenario calculations.

\begin{table}[ht]
\caption{Conditional geometric scenarios for $A_R=120$ cm$^2$, $\rho_t=1$ g/cm$^3$, $r=3$, $\Delta T_v=1$ K, $N=5.02\W$, $C=0.05\W$, $\gamma=0.04$ and $s=0.0675$. The values are not human flow estimates.}
\label{tab:S-mass}
\centering\small
\begin{tabular}{@{}rrrrrr@{}}
\toprule
$d_R$ (mm)&$M_R$ (g)&$\dot V_n$ (mL/min)&$H$ (W)&$\gamma H$ (W)&$\Q$ (W)\\
\midrule
0.5&6&5.94&0.378&0.015&0.400\\
1.0&12&11.88&0.756&0.030&0.414\\
2.0&24&23.76&1.511&0.060&0.442\\
3.0&36&35.64&2.267&0.091&0.470\\
4.0&48&47.52&3.022&0.121&0.498\\
4.21&50.5&50.0&3.180&0.127&0.504\\
\bottomrule
\end{tabular}
\end{table}

For $\gamma>0$ and $rM_R>0$, the reduced allocation requires
\begin{equation}
 \Delta T_v\le\frac{N-C}{0.020988\,\gamma rM_R}\ \mathrm{K}.
 \label{eq:S-weak}
\end{equation}
If a closed local energy interpretation also requires $C+H\le N$, it gives the stronger ceiling
\begin{equation}
 \Delta T_v\le\frac{N-C}{0.020988\,rM_R}\ \mathrm{K}.
 \label{eq:S-strong}
\end{equation}
The stronger inequality uses additional physical premises. The weaker one becomes arbitrarily loose as $\gamma\downarrow0$. Neither imposes a positive lower bound on venous cooling.

\subsection{A reference scenario}
Retaining the working inputs $N=5.02\W$, $C=0.05\W$, $H=3.18\W$, $\alpha=\beta=0.20$, $f=0.15$ and $\eta=0.45$ gives
\begin{align}
 \gamma&=0.04,& s&=0.0675,\\
 \gamma H&=0.1272\W,& N-C-\gamma H&=4.8428\W,\\
 s(N-C-\gamma H)&=0.326889\W,&\Q&=0.504089\W.
 \label{eq:S-legacy}
\end{align}
The systemic term supplies about $64.85\%$ of this output. The number is therefore strongly dependent on an unmeasured persistence fraction. Table~\ref{tab:S-eta} varies that fraction while retaining the other assumptions.

\begin{table}[ht]
\caption{Sensitivity of the same constructed reference scenario to the assumed systemic persistence fraction. These rows are not observations.}
\label{tab:S-eta}
\centering\small
\begin{tabular}{@{}rrrr@{}}
\toprule
$\eta$&Regional (W)&Systemic (W)&Total $\Q$ (W)\\
\midrule
0.00&0.1272&0.0000&0.1772\\
0.10&0.1272&0.0726&0.2498\\
0.20&0.1272&0.1453&0.3225\\
0.45&0.1272&0.3269&0.5041\\
0.80&0.1272&0.5811&0.7583\\
1.00&0.1272&0.7264&0.9036\\
\bottomrule
\end{tabular}
\end{table}

\begin{figure}[ht]
\centering
\includegraphics[width=0.82\linewidth]{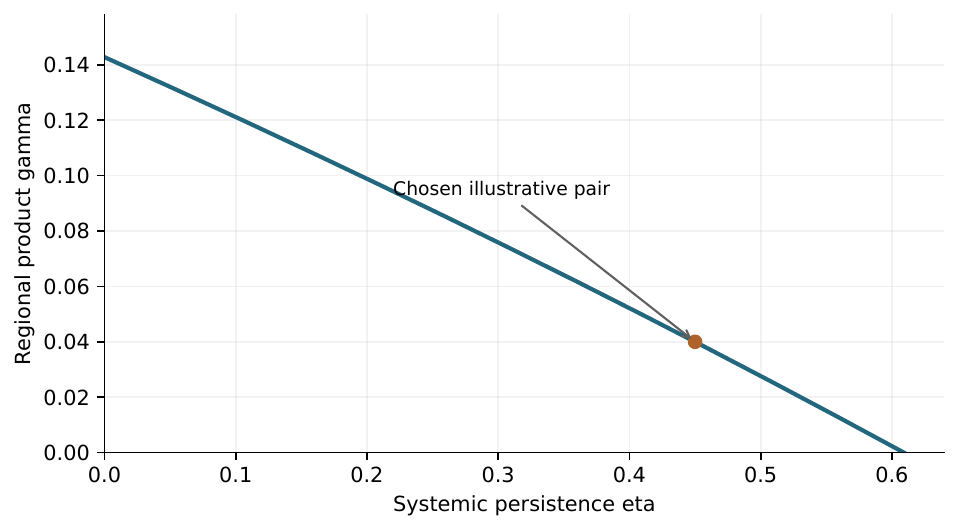}
\caption{Conditional single-endpoint identification curve for the reference hypothetical input values in \eqref{eq:S-legacy}. Each point yields the same prescribed $\Q$; the marked pair is the originally chosen illustration. All input powers and couplings in this example are imposed, not fitted.}
\label{fig:S-single}
\end{figure}

The sharp output interval for these imposed $N,C,H,f$ is $[0.05,3.4985]\W$. If the value $0.504089\W$ were hypothetically supplied as an exact observed endpoint, the main inverse formula would give
\[
 s\in\left[0,\frac{454089}{4970000}\right],\qquad
 \eta\in\left[0,\frac{151363}{248500}\right],\qquad
 \gamma\in\left[0,\frac{151363}{1060000}\right].
\]
Their upper endpoints are approximately $0.091366$, $0.609107$ and $0.142795$, respectively. The last two ranges are projections of a coupled curve, not independent intervals. A purely regional allocation with $s=0$ and an allocation with $\gamma=0$ both produce this endpoint under the imposed source scales (Fig.~\ref{fig:S-single}). This illustrates conditional ambiguity rather than measuring it in human subjects.

\section{Study-specific observables and numerical translations}
\subsection{Six source studies}
Table~\ref{tab:S-studies} records why selected studies do not jointly constitute a matched dataset for the inverse model. Reported variability retains its original role; a dispersion of local observations does not become uncertainty in an unmeasured whole-system coupling.

\begin{table}[p]
\caption{Distinct source studies, regimes and observables. The values are attributed to their original protocols and are not pooled.}
\label{tab:S-studies}
\small
\begin{tabularx}{\linewidth}{@{}>{\raggedright\arraybackslash}p{0.20\linewidth}YY@{}}
\toprule
Source & Regime and observation & Missing attribution information\\
\midrule
\citet{Bende1983} & $^{133}$Xe mucosal clearance in 80 humans; $33\pm10$ mL/min per $100$ g, mean $\pm$ SD. & Local mass-normalized perfusion does not supply total regional flow, deep venous temperature or cerebral power.\\[0.6em]
\citet{White1995} & Passive hyperthermia in five men; approximately threefold change in a local septal laser-Doppler signal. & A relative local signal is not a measured multiplicative change in whole-nose absolute flow.\\[0.6em]
\citet{Rouadi1999} & Imposed cold--dry and warm--humid airflow at 5, 10 and 20 L/min; nasopharyngeal conditioning and water gain. & One-way apparatus excludes expired-air contamination and does not measure net bidirectional-cycle export.\\[0.6em]
\citet{Na2020} & Numerical expiration model; scenario-dependent heat and water recovery. & Simulated recovery fractions do not calibrate net human nasal export or cerebral transfer.\\[0.6em]
\citet{Harris2007} & Factorial crossover in 12 brain-injured patients; nasal-air cerebral and oesophageal reductions of $0.15$ and $0.13\,^{\circ}$C. & The descriptive difference is not a significant selective effect or a measured heat rate.\\[0.6em]
\citet{Mariak1999} & Subdural point measurements in four postoperative patients; nasal inspiration/oral expiration during a three-minute maneuver. & Local response, cooled volume, nasal source and regional venous deficit were not jointly measured.\\
\bottomrule
\end{tabularx}
\end{table}

\subsection{Conversion of imposed-flow water-gain data}
For the cold--dry protocol of \citet{Rouadi1999}, the reported water gains over 15 minutes are converted to latent power using $2.425$ J/mg:
\begin{equation}
 P_{\mathrm{lat}}=\frac{2.425\,M_{15}}{900}\ \mathrm{W},
 \label{eq:S-Rouadi}
\end{equation}
with $M_{15}$ in mg. Table~\ref{tab:S-Rouadi} applies the same linear conversion to the reported errors.
\begin{table}[ht]
\caption{Latent-power translation of one-way water-gain measurements. The error terms are the reported standard errors propagated through the unit conversion, not uncertainty in net-cycle nasal export.}
\label{tab:S-Rouadi}
\centering\small
\begin{tabular}{@{}rrrr@{}}
\toprule
Flow (L/min)&Water gain (mg/15 min)&Rate (mg/min)&Latent power (W)\\
\midrule
5&$297.6\pm12.6$&19.84&$0.802\pm0.034$\\
10&$509.1\pm32.0$&33.94&$1.372\pm0.086$\\
20&$794.1\pm62.1$&52.94&$2.140\pm0.167$\\
\bottomrule
\end{tabular}
\end{table}
These powers characterize that imposed-flow water gain. They neither replace the illustrative $N=5.02\W$ nor confirm its assumed full-cycle nasal attribution. Applying an expiratory recovery fraction from a different simulation would require matched geometry, boundary conditions and flow-duty information.

\subsection{Positive cooling interventions}
\citet{Covaciu2011} studied ten awake volunteers with circulating-water nasal balloons; nine were evaluable for MR thermometry. The reported brain reductions after 60 minutes were approximately $1.7$--$1.8\,^{\circ}$C, with rectal measurements available in a smaller subset. \citet{AbouChebl2011} reported mean brain and core decreases of approximately $1.4$ and $1.1\,^{\circ}$C during active transnasal evaporative treatment in 15 brain-injured patients. These are observations under active interventions; neither temperature decrement is converted here into a natural nasal heat-transfer rate.

\citet{Nybo2014} used arterial and internal jugular measurements in six healthy subjects and found parallel cooling during the balloon intervention without a selective cerebral thermal advantage. The measured endpoints differ from the MR and local parenchymal endpoints above. Likewise, the high-flow swine experiment of \citet{Stonko2023}, transnasal ambient-air experiments in piglets \citep{Koehler2023}, and combined transnasal cooling and endovascular warming in swine \citep{Nishikimi2026} involve distinct imposed conditions. In the latter case, the systemic warming is part of the intervention. These studies support investigation of transnasal thermal interventions without identifying the passive human coefficients in \eqref{eq:S-model}.

\section{An idealized transient conduction construction}
\subsection{Pulse solution and maximum}
Let $u(x,t)$ be a positive temperature decrement in a homogeneous, nonperfused half-space $x>0$, with thermal diffusivity $\kappa>0$. Prescribe $u(x,0)=0$, $u(0,t)=\Delta T_m$ for $0<t<\tau$, and $u(0,t)=0$ after the pulse. The diffusion equation $u_t=\kappa u_{xx}$ gives, at depth $L>0$,
\begin{equation}
 \frac{u(L,t)}{\Delta T_m}=
 \begin{cases}
 \displaystyle\operatorname{erfc}\!\left(\frac{L}{2\sqrt{\kappa t}}\right),&0<t\le\tau,\\[0.9em]
 \displaystyle\operatorname{erfc}\!\left(\frac{L}{2\sqrt{\kappa t}}\right)
 -\operatorname{erfc}\!\left(\frac{L}{2\sqrt{\kappa(t-\tau)}}\right),&t>\tau.
 \end{cases}
 \label{eq:S-pulse}
\end{equation}
The second line follows by superposing a positive boundary step at zero and a negative step at $\tau$.

\begin{proposition}[Unique delayed maximum]
For $L,\kappa,\tau,\Delta T_m>0$, the response in \eqref{eq:S-pulse} has a unique global maximum at $t_*>\tau$, determined by
\begin{equation}
 \frac{L^2}{4\kappa}
 =\frac{3t_*(t_*-\tau)}{2\tau}
 \log\!\left(\frac{t_*}{t_*-\tau}\right).
 \label{eq:S-peak}
\end{equation}
\end{proposition}
\begin{proof}
The first line of \eqref{eq:S-pulse} is strictly increasing in $t$. Put $a=L^2/(4\kappa)>0$. For $t>\tau$, differentiation shows that stationarity is equivalent to
\[
 t^{-3/2}e^{-a/t}=(t-\tau)^{-3/2}e^{-a/(t-\tau)}.
\]
Taking logarithms gives \eqref{eq:S-peak}. With $x=(t-\tau)/\tau>0$, its right side is $(3\tau/2)x(x+1)\log(1+1/x)$. The derivative has the sign of
\[
 (2x+1)\log(1+1/x)-1>\frac{2x+1}{x+1}-1>0.
\]
The right side of \eqref{eq:S-peak} tends to zero at $t\downarrow\tau$ and diverges as $t\to\infty$, so there is exactly one stationary time. The response is increasing just after $\tau$, positive for finite $t>0$, and tends to zero at infinity. The stationary point is therefore its unique global maximum.
\end{proof}

\begin{figure}[ht]
\centering
\includegraphics[width=0.91\linewidth]{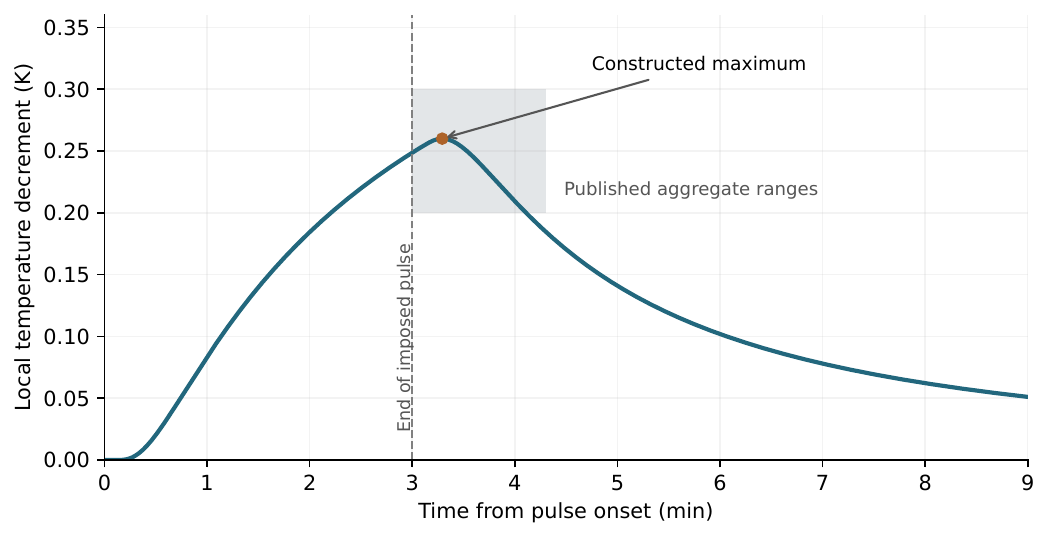}
\caption{Constructed conduction-only response. The rectangle marks the published aggregate time and amplitude ranges for the maneuver in \citep{Mariak1999}; it is not a confidence region or digitized patient trajectory. The surface temperature and effective depth are chosen inputs.}
\label{fig:S-conduction}
\end{figure}

Choose $k=0.50$ W/(m K) and $c_{\mathrm{vol}}=3.885\times10^6$ J/(m$^3$ K), so $\kappa=k/c_{\mathrm{vol}}\simeq1.287\times10^{-7}$ m$^2$/s, a scale used in bioheat calculations \citep{ZhuDiao2001}. With $L=6$ mm and $\tau=180$ s, the computed maximum occurs at $197.545$ s. Its normalized height is approximately $0.39536$; a prescribed surface decrement $\Delta T_m\simeq0.65763$ K gives a peak of $0.26$ K (Fig.~\ref{fig:S-conduction}).

For comparison, \citet{Mariak1999} reported additional local decrements of $0.20$--$0.30\,^{\circ}$C and minimum times of $3.0$--$4.3$ min from the start of an intensive three-minute breathing maneuver. The construction lies within those two aggregate ranges. However, $L$ and the surface decrement are free, unmeasured choices here. The half-space omits perfusion, bone layers and the actual skull-base geometry. Matching two summaries in that idealized class does not establish compatibility with all patient anatomy or show that conduction caused the observed response.

\subsection{Energy stored in the transient deficit}
For an effective boundary area $A_c$, integration over the half-space gives the energy deficit relative to its initial state:
\begin{equation}
 E(t)=\frac{2kA_c\Delta T_m}{\sqrt{\pi\kappa}}
 \begin{cases}
 \sqrt t,&0<t\le\tau,\\
 \sqrt t-\sqrt{t-\tau},&t>\tau.
 \end{cases}
 \label{eq:S-energy}
\end{equation}
This follows from $\int_0^\infty\operatorname{erfc}(x/(2\sqrt{\kappa t}))\,\dd x=2\sqrt{\kappa t/\pi}$ and $k=c_{\mathrm{vol}}\kappa$. At pulse termination,
\begin{equation}
 E(\tau)=2kA_c\Delta T_m\sqrt{\frac{\tau}{\pi\kappa}}.
\end{equation}
With the additionally assumed $A_c=5$ cm$^2$, this is approximately $6.94$ J, or $0.0385\W$ averaged over the imposed pulse. After the boundary is restored, heat enters the medium and $E(t)\to0$. Thus the deficit at $\tau$ is not permanent net extraction over the entire relaxation and is not the steady conductive allocation $C$ or the cycle-averaged nasal budget $N$. The relevant boundary area and temperature were not measured in the cited experiment.

\section{Feasible polygons and coefficient bounds}
\subsection{Sharp endpoint-error compatibility}
For the known-input repeated-condition problem, define $a_i=N_i-C_i$, $y_i=q_i-C_i$ and $t=(1-s)\gamma$. For fixed $0<f<1$, the exact domain is
\begin{equation}
 \mathcal P=\{(s,t):0\le s\le f,\ t\ge0,\ s+t\le1,\ a_i s+H_i t\le a_i\ \forall i\}.
\end{equation}
For endpoint intervals $[\ell_i,u_i]$, intersect $\mathcal P$ with $\ell_i\le a_i s+H_i t\le u_i$. Every point of the resulting set yields a compatible original coefficient pair via $\gamma=t/(1-s)$ and $\eta=s/f$, and every compatible original pair maps back to this set. The construction is therefore sharp conditional on the input matrix and the endpoint bands.

Linear projections such as those onto $s$ and $t$ attain extrema at vertices. The same is true for $\gamma=t/(1-s)$ on any nonempty feasible polygon, including a segment or singleton. To see this, write its vertices as $(s_j,t_j)$ and any convex representation as $(s,t)=\sum_j\lambda_j(s_j,t_j)$. Then
\[
 \frac{t}{1-s}
 =\sum_j\frac{\lambda_j(1-s_j)}{\sum_k\lambda_k(1-s_k)}
 \frac{t_j}{1-s_j}.
\]
The displayed weights are nonnegative and sum to one, since $s_j\le f<1$. Thus $\gamma$ lies between its smallest and largest vertex values, and both bounds are attained.

For the main synthetic example with endpoint errors bounded by $0.01\W$, the four vertices are
\[
 (0.06,0.1228),\quad (0.05,0.1428),\quad
 (0.06,0.1028),\quad (0.07,0.0828).
\]
The corresponding projection bounds are
\[
 0.05\le s\le0.07,\quad
 0.0828\le t\le0.1428,\quad
 \frac{69}{775}\le\gamma\le\frac{357}{2375},
\]
with regional endpoints approximately $0.08903226$ and $0.15031579$, and $1/3\le\eta\le7/15$ for the fixed $f=0.15$. These are deterministic conditional sets. They are not confidence intervals and they do not include uncertainty in the source, reference, conductive allocation, regional scale or coupling invariance.

\subsection{Independent consistency conditions}
A third known input row can test a coefficient pair identified from the first two. For example, adding $C_3=0.03\W$, $a_3=3\W$ and $H_3=2\W$ to the main synthetic coefficients predicts
\[
 y_3=(0.06)(3)+(0.1128)(2)=0.4056\W,\qquad q_3=0.4356\W.
\]
An exact different value would contradict the combined fixed-input and common-coefficient assumptions. With endpoint tolerances, all three bands must be intersected. Failure is not by itself evidence against one specific anatomical mechanism; the source reconstruction, endpoint definition or invariance assumptions may also fail.

\section{Local sensitivity and statistical qualifications}
\subsection{Derivatives within the feasible domain}
Using $\Q=C+f\eta(N-C)+(1-f\eta)\gamma H$, the derivatives are
\begin{align}
 \partial_N\Q&=f\eta,&\partial_C\Q&=1-f\eta,\\
 \partial_H\Q&=(1-f\eta)\gamma,&\partial_f\Q&=\eta(N-C-\gamma H),\\
 \partial_\gamma\Q&=(1-f\eta)H,&\partial_\eta\Q&=f(N-C-\gamma H).
 \label{eq:S-derivatives}
\end{align}
Since $\gamma=\alpha\beta$, $\partial_\alpha\Q=(1-f\eta)\beta H$ and $\partial_\beta\Q=(1-f\eta)\alpha H$. The mixed derivative
\begin{equation}
 \frac{\partial^2\Q}{\partial\gamma\,\partial\eta}=-fH
\end{equation}
expresses the allocation of a shared residual; it does not demonstrate physiological inhibition between pathways.

These derivatives describe independent mathematical perturbations. At an active constraint such as $C+\gamma H=N$, some coordinate directions leave the feasible domain. A physical sensitivity must respect feasible directions and any dependence among the inputs. An elasticity $(z/\Q)\partial_z\Q$ can be interpreted as relative sensitivity only where the relevant parameter and $\Q$ are positive.

\subsection{First-order variance propagation}
If a justified, sufficiently concentrated joint statistical distribution exists for the input vector $\theta$, with covariance $\Sigma$, the delta-method approximation is
\begin{equation}
 \operatorname{Var}(\Q)\simeq\nabla\Q^{\mathsf T}\Sigma\nabla\Q.
 \label{eq:S-delta}
\end{equation}
This is a local statistical approximation. It does not assign a distribution to unmeasured coupling fractions and must not use between-subject scatter from a different observable as though it were uncertainty in the present coefficients.

For $\gamma=\alpha\beta$, a first-order approximation is
\begin{equation}
 \operatorname{Var}(\gamma)\simeq
 \mu_\beta^2\operatorname{Var}(\alpha)
 +\mu_\alpha^2\operatorname{Var}(\beta)
 +2\mu_\alpha\mu_\beta\operatorname{Cov}(\alpha,\beta).
\end{equation}
For independent factors with finite second moments, the exact expression additionally includes the product of variances:
\begin{equation}
 \operatorname{Var}(\alpha\beta)=
 \mu_\beta^2\operatorname{Var}(\alpha)
 +\mu_\alpha^2\operatorname{Var}(\beta)
 +\operatorname{Var}(\alpha)\operatorname{Var}(\beta).
\end{equation}
No probabilistic intervals for human coupling coefficients are produced in this study. The main text instead reports deterministic feasibility and explicitly fixed-input endpoint bands.

\section{Reproduction and numerical verification}
The source archive contains \texttt{analysis/reproduce\_revision.py} for the steady and conduction examples and \texttt{analysis/verify\_extensions.py} for the variable-flow and storage extensions. Running the scripts with Python, NumPy, SciPy, SymPy and Matplotlib regenerates all four figures, scenario CSV files and verification summaries. The pre-generated vector PDF figures allow Overleaf to compile the manuscript without Python.

The numerical checks compare the analytical output and single-endpoint bounds with independently formulated linear programs over the transformed feasible polygon, verify the synthetic inversion and boundary cases, and reproduce the transient maximum, energy and source-unit conversions. They check the calculations under the stated equations. They cannot validate physiological source attribution, coefficient invariance or a human cooling effect.

\section{Connection with a compartmental bioheat balance}
\label{sec:S-nodes}
Classical bioheat theory \citep{Pennes1948} and individualized whole-body models \citep{Takahashi2021} provide context for constructing a forward thermal model. A schematic well-mixed tissue-node balance is
\begin{equation}
 \mathcal C_j\frac{\dd T_j}{\dd t}
 =\sum_k K_{jk}(T_k-T_j)
 +\rho_b c_{p,b}\dot V_j(T_{a,j}-T_j)
 +P_{\mathrm{met},j}+P_{\mathrm{ext},j}.
 \label{eq:S-node}
\end{equation}
Here $\mathcal C_j$ is heat capacity in J/K, $K_{jk}$ is conductance in W/K and $\dot V_j$ is perfusion in m$^3$/s; all right-hand terms are powers. This equation is a generic proposed balance and is not a transcription of JOS-3. Its blood-exchange term assumes an effective common tissue temperature and a specified arterial inlet temperature. A closed network must include the opposite exchange term in its blood balance. For symmetric $K_{jk}$, paired internal conduction terms cancel on summing over nodes. A prescribed arterial reservoir alone does not close the whole-body energy budget.

The documentation of JOS-3 includes respiratory heat loss, cardiac output and segmental temperatures \citep{JOS3Repository}. These outputs have distinct spatial scopes and units. Whole-respiratory heat loss cannot directly supply a nasal-specific source; a modeled head-core flow is not automatically an independently measured cerebral flow. If an existing respiratory sink includes nasal exchange, any resolved nasal component must be introduced through an explicit partition of that sink. Adding both contributions would count the same exchange twice.

The source ledger in the main text is an attribution model at a different level of description. To connect it to \eqref{eq:S-node}, one must define the nasal boundaries and common reference, derive the corresponding source and endpoint from the network, and test whether constant effective couplings describe those derived quantities. A synthetic reconstruction of the simulator's own parameters tests the inverse calculations, not the physiological validity of the simulator or the source attribution.

\section{Transient rate identification and reproducible examples}
\label{sec:S-transient}
Write the main storage extension as
\[
 E'=R-\sigma E,\quad q_{\mathrm{dyn}}=B+fk_sE,
 \quad R=N-B\ge0,\quad \sigma=k_s+k_r>0.
\]
Suppose the source functions $N,B$ and a measurable $f$ with $0<f_{\min}\le f(t)\le1$ are known. Suppose also that the reconstructed trajectory gives an absolutely continuous, nonnegative
\[
 w(t)=\frac{q_{\mathrm{dyn}}(t)-B(t)}{f(t)}.
\]
This hypothesis holds for compatible model outputs, since $w=k_sE$. The division uses the known flow fraction; no differentiation of $f$ is required.

\begin{proposition}[Identification of constant storage rates]
\label{prop:S-rates}
For intervals $I_j=[u_j,v_j]$, define the known integrals and increments
\[
 A_j=\int_{u_j}^{v_j}R(t)\,\dd t,\qquad
 W_j=\int_{u_j}^{v_j}w(t)\,\dd t,\qquad
 d_j=w(v_j)-w(u_j).
\]
Every compatible constant rate pair satisfies
\begin{equation}
 \begin{pmatrix}A_1&-W_1\\\vdots&\vdots\\A_m&-W_m\end{pmatrix}
 \begin{pmatrix}k_s\\\sigma\end{pmatrix}
 =\begin{pmatrix}d_1\\\vdots\\d_m\end{pmatrix},
 \qquad 0\le k_s\le\sigma,\quad \sigma>0.
 \label{eq:S-rate-linear}
\end{equation}
If the displayed matrix has rank two, there is at most one compatible pair, determining $k_r=\sigma-k_s$. A candidate obtained from finitely many intervals must additionally satisfy
\begin{equation}
 w(t)-w(0)=k_s\int_0^tR(r)\,\dd r-\sigma\int_0^tw(r)\,\dd r
 \quad\text{for every }t
 \label{eq:S-rate-full}
\end{equation}
and the initial-state condition. For $k_s>0$, these conditions are sufficient with $E=w/k_s$ and $E_0=w(0)/k_s$; if $E_0$ is prescribed independently, its value must agree.
\end{proposition}
\begin{proof}
For compatible data, $w=k_sE$ and hence $w'=k_sR-\sigma w$ almost everywhere. Integration gives \eqref{eq:S-rate-linear} and \eqref{eq:S-rate-full}. Rank two makes the kernel zero, proving uniqueness of a compatible pair. Conversely, for $k_s>0$, absolute continuity and \eqref{eq:S-rate-full} give $E'=R-\sigma E$ for $E=w/k_s\ge0$. The definition of $w$ reconstructs the output, and $k_r=\sigma-k_s\ge0$ gives the prescribed model. The initial condition follows as stated.
\end{proof}

At $k_s=0$, compatibility requires $w\equiv0$. The endpoint then contains no information about $k_r$ or $E_0$. A hypothetical nonzero $w$ satisfying a homogeneous decay equation is not a compatible $k_s=0$ output. More generally, constant stationary data satisfy $w=(k_s/\sigma)R$, so all nonzero interval rows are proportional and only the rate ratio is supplied. These distinctions explain why a transient is required to identify a time scale.

\subsection{Loading and source-free decay}
Take the synthetic main-text example: $R=0.8\W$ on $[0,60]$ s, $R=0$ afterwards, $k_s=0.02$ s$^{-1}$, $k_r=0.03$ s$^{-1}$ and $E_0=0$. Then
\[
 E(t)=\begin{cases}
 16(1-e^{-0.05t})\ \mathrm J,&0\le t\le60,\\
 16(1-e^{-3})e^{-0.05(t-60)}\ \mathrm J,&t>60.
 \end{cases}
\]
On the loading interval $A_1=48$ J; on the subsequent decay interval $[60,120]$ s, $A_2=0$ and $W_2>0$. The determinant of the two-row matrix is $-48W_2\ne0$. Consequently the exact integrals recover $k_s=0.02$ s$^{-1}$ and $\sigma=0.05$ s$^{-1}$. A source-free positive tail also determines $\sigma$ from its logarithmic decay, but it cannot determine $k_s$ without information on loading or the stored state.

At 180 s the total input is 60 J. The cerebral allocation is approximately 14.877739 J, the other allocation 45.084576 J and the remaining store 0.037685 J. Their sum is 60 J to floating-point precision. These quantities refer exclusively to the imposed synthetic example.

\subsection{Uncertainty and computational checks}
The interval formulation avoids numerically differentiating a reconstructed trajectory. It does not remove measurement error: uncertain $w$ affects both $d_j$ and $W_j$, while source uncertainty affects $A_j$. Treating only the increments as uncertain would neglect errors in the design matrix. No confidence interval for human storage rates is reported here.

The companion script \texttt{analysis/verify\_extensions.py} checks exact recovery for variable-flow designs, rejects an inconsistent lifted product, verifies a saturated-source degeneracy, and recovers the storage rates from the loading and decay integrals. It also compares the closed-form storage solution with independent numerical integration, including the accumulated energy balance. The resulting trajectories and summaries are included with the source project.